\documentclass[%
article,
superscriptaddress,
nofootinbib,
amsmath,amssymb,
aps,
pra,
notitlepage,
showkeys]{revtex4-2}

\usepackage[margin=3cm]{geometry}
\usepackage{color}
\usepackage{url}
\usepackage{graphicx}
\usepackage{lipsum}
\usepackage[export]{adjustbox}
\usepackage[font=footnotesize,labelfont=bf]{caption}
\usepackage{mathtools}
\usepackage{gensymb}
\usepackage[T1]{fontenc}
\usepackage{amsmath}
\usepackage{amssymb}
\usepackage{hyperref}
\usepackage{stmaryrd}
\usepackage[utf8]{inputenc}
\usepackage{rotating}
\usepackage{comment}
\usepackage{url}
\usepackage{bbm}
\usepackage{natbib}
\usepackage{soul}
\usepackage[normalem]{ulem}
\usepackage{tabularx,booktabs}
\usepackage{stmaryrd}
\usepackage{tcolorbox}
\usepackage{xcolor}
\usepackage{ulem}
\definecolor{myblue}{RGB}{119, 170, 218}
\definecolor{mygreen}{RGB}{44, 126, 51}

\newcolumntype{Y}{>{\centering\arraybackslash}X}

\begin{document}
\title{Three Centuries of the Laws of Cricket Reveal Core Principles of the Evolution of Regulatory Mechanisms}
\author{Daniel Chia}
\affiliation{Carnegie Mellon University, Pittsburgh, PA, USA.}

\author{Hyejin Youn}
\affiliation{Santa Fe Institute, Santa Fe, NM, USA.}
\affiliation{Graduate School of Business, Seoul National University, Seoul, South Korea.}

\author{Jonny Singer}
\affiliation{Marylebone Cricket Club, Lord's Cricket Ground, London, UK.}

\author{Dawoon Jeong}
\affiliation{Department of Sociology, University of Chicago, Chicago, IL, USA.}
\affiliation{Knowledge Lab, University of Chicago, Chicago, IL, USA.}

\author{Chris Kempes}
\affiliation{Santa Fe Institute, Santa Fe, NM, USA.}

\author{C.~Brandon Ogbunugafor}
\affiliation{Santa Fe Institute, Santa Fe, NM, USA.}
\affiliation{Department of Ecology and Evolutionary Biology, Yale University, New Haven, CT, USA.}
\affiliation{Department of Anthropology, Yale University, New Haven, CT, USA.}

\author{Geoffrey B.~West}
\affiliation{Santa Fe Institute, Santa Fe, NM, USA.}
\affiliation{Department of International Development, The London School of Economics and Political Science, London, UK.} 

\author{James Holehouse}
\email{jamesholehouse1@gmail.com}
\affiliation{Santa Fe Institute, Santa Fe, NM, USA.}
\affiliation{School of Biology, Washington University in St. Louis, St. Louis, Missouri, USA.}

\date{\today}

\begin{abstract}
    \vspace*{2em}
    \noindent Rules, regulations, and regulatory systems are central to societies, institutions, and organisms, yet surprisingly little is known about their evolution over long timescales. The Laws of Cricket, the world’s second most popular sport, offer a unique insight into this fundamental question. Their 268-year history constitutes the longest continuous rule-set record yet assembled. Our quantitative analysis reveals generic features including rule-book size growing exponentially in time but scaling sublinearly with matches played; new situations stimulate new rules, but at a decelerating rate; regulatory structures exhibit abrupt phase transitions, increasing rule specificity, interconnectivity and complexity with central rules shifting from gameplay to officiating. These provide a framework for understanding how governance evolves from simple collections of rules to complex regulatory architectures across social, legal, and biological domains.
\end{abstract}

\maketitle


\section*{Introduction}


\noindent Sports rules are archetypal rule systems. From an early age, sports provide many people with their first encounter with formal systems of rules, whether by inventing games of their own or by playing games governed by established regulations. While the importance of rules and regulations in human society is widely acknowledged \cite{daston2022rules}, relatively little is known about the evolution of regulatory mechanisms starting from simple lists of rules to the highly complex interconnected rule systems that govern society today. In this study, we explore the evolution of the Laws of Cricket—the world's second most popular spectator sport—across more than 268 years of written Laws. To our knowledge, this represents one of the longest continuous records of regulatory evolution available for quantitative study, spanning a longer period than comparable datasets from team sports, legal codes, and gene-regulatory systems. As a point of contrast, the original Laws of Cricket predate the Industrial Revolution by around 20 years, the US Constitution by 37 years, and the first edition of the United States Code (US Code) by 174 years. In short, the Laws of Cricket provide an unusually rich baseline dataset to understand the evolution of rules and regulations.

\begin{figure}[ht]

    \includegraphics[width=\textwidth]
    {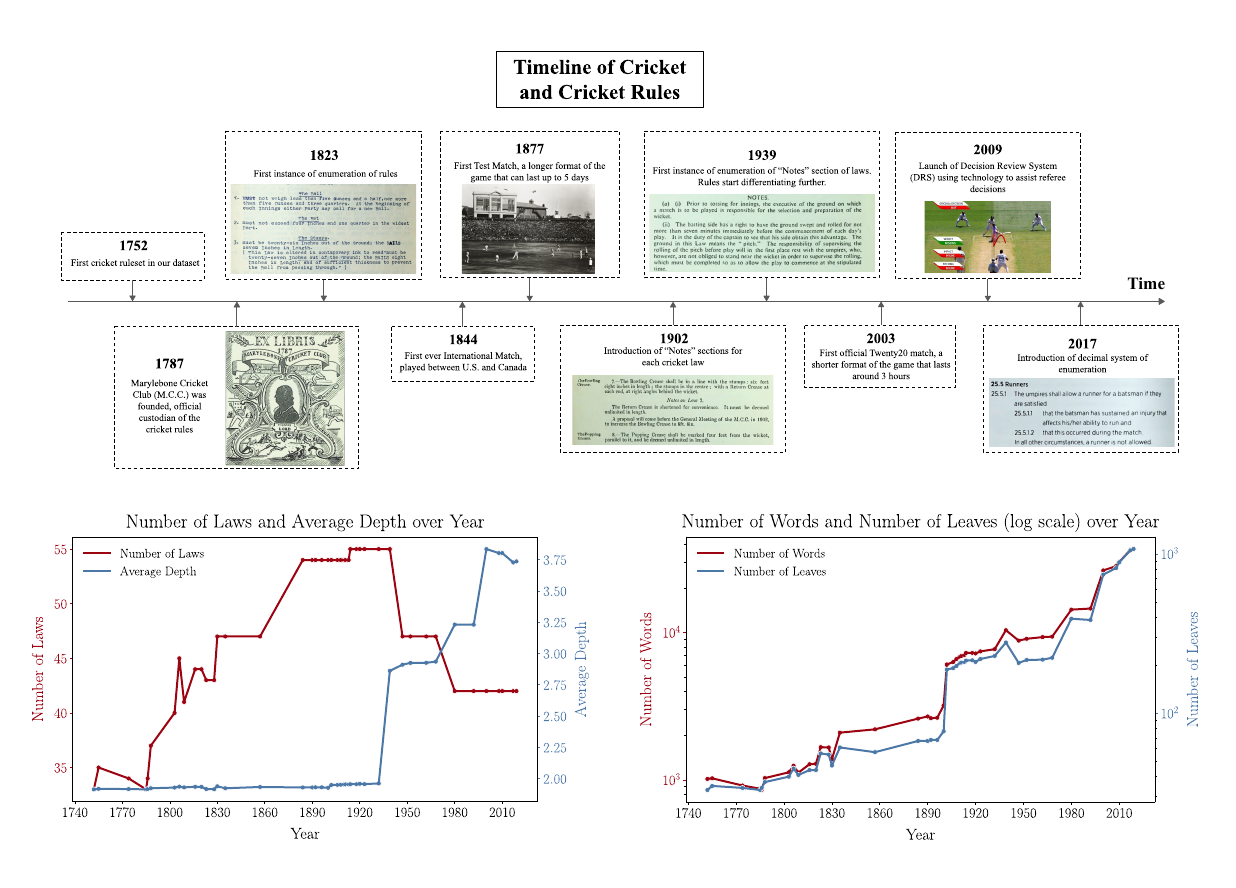}
    \caption{\textbf{Overview of landmark events and summary of long-run trends in cricket rule sets from the dataset.} For definitions of rules, Laws, depth, and leaves, please see the glossary in Box 1. \textbf{Top}: Timeline highlighting major developments in the game of cricket and key changes to its rules. \textbf{Bottom left}: Number of Laws (red) and average structural depth (blue), plotted against year. \textbf{Bottom right}: Total word count (red) and number of leaves (blue), also plotted against year on a logarithmic y-axis.
    \newline
    }
    \label{fig:timeline}
\end{figure}

From its inception to the present day, the Laws of Cricket have always been centered around promoting competitive, safe, and fair play, even though the cultural environment the sport exists in has changed substantially. That is, the fundamental purpose of the Laws of Cricket has not changed. Conveniently for the consistency of the dataset, the Laws of Cricket have for last 239 years been instituted by the same governing authority, the Marylebone Cricket Club. 
In Figure \ref{fig:timeline} we summarize the timeline of the Laws of Cricket, including the evolution of key rule set statistics in time (see Box 1 for definitions of key terms). Compared to other regulatory codes such as the US Code, a greater fraction of rule evolution in cricket is driven by endogenous processes within the system of the game and its rules: the effects that rule changes have on the game, and conversely the way that changes in gameplay compel certain adjustments to the rules. Sports arise naturally across all human cultures and civilizations, and for many sports their player base also spans national and cultural boundaries. This ubiquity makes sports a well-placed system for insight into the evolution of model regulatory systems. 

More broadly, regulation is ubiquitous across both social and biological systems. In social systems, regulations are encoded in codified lists of rules \cite{daston2022rules} and enforced by a diverse array of lawyers, tax officials, law enforcement, and judges. In contrast, gene-regulatory networks implement regulation via circuits of interacting genes \cite{babu2004structure}, facilitated by transcription factors, regulatory RNAs, and regulatory architecture such as chromatin and insulator proteins \cite{schultz2026topological}. In short, codified rules and gene-regulatory circuits constitute regulatory architectures that constrain and coordinate social and biological processes, respectively. 
In turn, social environments create pressures on the social processes that occur within them, thus driving change in their codified rules; ecologies create pressures on the organisms within them, driving change in their gene circuits \cite{babu2004structure,heinrich2012regulation,katz2020complex,march2000dynamics,daston2022rules,davidson2010emerging,erwin2009evolution}. Thus, regulatory evolution is driven by a feedback loop between a system, its external environment  and the regulatory architecture through which it is governed.


\begin{center}
\begin{tcolorbox}[width=1.0\textwidth,title={\small \textbf{Box 1: Glossary}}] 
   \small{
    \textbf{Rule book}: The set of laws/rules comprising the game of cricket for a given year. We also refer to the rule book as the \textit{Laws of Cricket} for a given year (the terminology used by the rule makers) and the \textit{rule set}.\\
    \textbf{Rule}: A paragraph of text from a cricket rule book that corresponds to a leaf in the tree diagram representing rule book structure. Rule makers refer to this as a \textit{clause}, and we may also refer to rules as \textit{leaves} in the context of the rule tree. See Figure \ref{fig:tree_nomenclature} for more details.\\
    \textbf{Law}: The highest-order category in the rule book. Laws categorize the most important distinctions in the rule book. ``Law'' by convention is always capitalized. See Figure \ref{fig:tree_nomenclature} for more details.\\
    \textbf{Rule book structure}: The tree structure containing the laws, structural nodes, and rules for a rule set in a given year. For examples, please see Figures \ref{fig:tree_nomenclature} and \ref{fig:tree_evolution}. We also refer to the rule book structures as \textit{rule trees} given their inherent tree structures.\\
    \textbf{Rule depth}: The number of nested levels a rule inhabits in the rule book. For example, Rule 2.3.1.4 has depth 4 (root nodes of rule book year has depth 0). \\
    \textbf{Cross reference}: When a rule cites the information contained in another rule. For examples, please see scanned images of the rule books in Figure \ref{fig:tree_evolution}.\\
    \textbf{Interdependency network}: The weighted and directed network formed for all the the cross references of rules in a given year, coarse grained to the level of Laws. For examples please see the top row of Figure \ref{fig:citation_network_evolution}.
    }
\end{tcolorbox} 
\end{center}

The roadblocks to constructing universal frameworks of rule evolution range from the theoretical to the methodological. The first set of roadblocks are theoretical. There is a tendency to study individual rule systems in silos, instead of viewing them as a collective body of rule systems that might share structural universalities. For example, individual studies have looked at the evolution of internet protocol \cite{medina2005measuring}, rules in Stanford University \cite{march2000dynamics}, sports rules \cite{Vamplew01072007}, and labor governance rules \cite{hassel2008evolution}. Because of the emphasis on drawing findings about specific systems rather than the dynamics of rules themselves, these studies are more qualitative than quantitative in nature. It is worth noting a couple of exceptions: (i) on the societal level, Katz and colleagues conducted quantitative analyses of the US Code \cite{coupette2021measuring,katz2020complex,bommarito2017measuring}; (ii) on the biological level there is work on the scaling of regulatory genes with respect to genome size \cite{van2003scaling}---scaling principles that have recently been extended to generic bureaucracy scaling and growth \cite{yang2024regulatory,yang2024leads}. 

The second set of roadblocks lie in the \textit{lack of rule set data presenting the complete evolution of a regulatory system from simple beginnings}. While many rule sets exist in hardcopy form, it is usually only the more recent versions of rule sets that exist in softcopy form and are thus readily amenable to computational analysis. Additionally, across social systems, focus is often on the state of the law as it currently stands, or else case studies of how particular aspects of the law change in time. Generally, the focus is not on how the structures across all of the law changes through time (with refs.~\cite{katz2020complex,jeong2026dataset} being of the few exceptions to this). Herein, we address this issue by constructing a machine-readable version of the Laws of Cricket through the centuries that was previously only available in hardcopy---constituting one of longest timespan rule sets digitally recorded (268 years). By constructing trees of rule book structures, we can define rules consistently as nodes in a tree, a definition independent of semantics and which extends across other unrelated rule sets such as the US Code.


This article explores four key theoretical questions using longitudinal data from the Laws of Cricket: (1) \textit{How does rule book size scale with the exposure to novel scenarios?} While analogous questions have been addressed for some biological and social systems \cite{van2003scaling,yang2024regulatory}, the mechanistic-scaling of rule sets remains relatively unexplored. (2) \textit{How is the specificity, or depth, of individual rules changing over time?} Building on qualitative distinctions between ``thick'' and ``thin'' rules \cite{daston2022rules}, we develop metrics to quantify the specificity of rules in a rule set using trees. (3) \textit{How does rule-book structure change over time?} Using rule trees as models of hierarchical organization, we identify phase transitions in the structure of the Laws of Cricket. (4) \textit{Do rules become increasingly interdependent as rule books evolve?} We show that individual laws in the game of cricket are becoming increasingly interdependent, and that the laws considered to be the most central to the game have changed qualitatively in time: from laws centered around gameplay to laws more centered around the \textit{spirit of the game}. In summary, our study uses the extensive dataset of the Laws of Cricket to provide a quantitative analysis of how the complexity of rules evolves over time, shedding light on the principles of regulatory evolution.

\section*{Results}
\noindent The results that follow examine multiple dimensions of rule-book evolution and are therefore necessarily technical in places. To aid readers primarily interested in the broader implications of the study, we begin with a summary of the principal findings, i.e., the core principles of the evolution of regulatory mechanisms:
\begin{enumerate}
    \item While rules grow exponentially in time, the size of the rule books---measured in terms of \textit{number of rules} or \textit{words in the rule book}---scales as a sublinear power-law with the number of matches that have been played. This implies that as more matches are played the rate of rule addition per game decreases. This indicates that discovering novel ways to break the rules of the game becomes more difficult as rule books evolve. See Figure \ref{fig:textual_analysis}.
    \item While the size of the rule books is a relatively smooth function of the number of matches played (and time), the structures inherent to the rules have \textit{phase transitions} in time. These phase transitions represent reorganizations of the rule sets into deeper and more regular structures. The process of reorganization is driven by the recategorization of important coarse-grained aspects of the game (Laws) and the branching of individual rules into more fine-grained categories. Necessarily, individual rules admit \textit{thinner} as opposed to \textit{thicker} characteristics (using the language of ref.~\cite{daston2022rules}). See Figures \ref{fig:tree_evolution} and \ref{fig:tree_analysis}.
    \item As rule books grow in size, individual rules become increasingly interdependent and cite one another. Individual rules then become dependent on other rules, and individual aspects of gameplay become dependent on multiple connected rules. This indicates that rule book evolution naturally leads to more pleiotropy in the relationship between the sport as played (akin to a phenotype), and the rules in the rule book (akin to a genotype). See Figure \ref{fig:cit-ntwk-stat-summ}.
    \item The themes of the most central Laws in the rule books exhibit a systematic shift in time. At the initial stages of rule interdependence, the most central Laws of the game were related to gameplay. However, as the rule sets evolve, the most important Laws increasingly center around aspects of the game related to officiating. This indicates a general trend for increasingly central rules to concern the governance and administration of the game itself. See Figures \ref{fig:top-five-ev-centrality} and \ref{fig:three-case-studies}.
\end{enumerate}
Readers primarily interested in the broader implications of these findings should proceed directly to the Discussion, while readers interested in the details of the analyses should continue below.

\subsection*{Text-based measures of evolution}


\noindent Fundamentally, the rules and laws of social systems are in the semantics of passages that make up the rule books. Although in later sections we focus on the structures of the rules in the rule books, an important starting point looks at the basic word-statistics and semantics of the Laws of Cricket, which can be thought of as the genetic code governing how the sport is played. Text-based measures have been used extensively in computational analysis to understand the semantic, syntactic and lexical patterns of texts. For example, Zipf-Mandelbrot fits have been widely observed to approximate the rank-frequency distributions across a multitude of languages \cite{zipf_psycho-biology_1935, mandelbrot1953informational, balasubrahmanyan1996, Ferrer_i_Cancho2005-se}, and their parameters have been used to capture shifts in morphological structure in languages \cite{Bentz2014-wz} and to measure lexical diversity and standardization in scholarly titles \cite{berube2018lexdiv}.

\begin{figure}[ht]
    \includegraphics[width=0.8\textwidth]{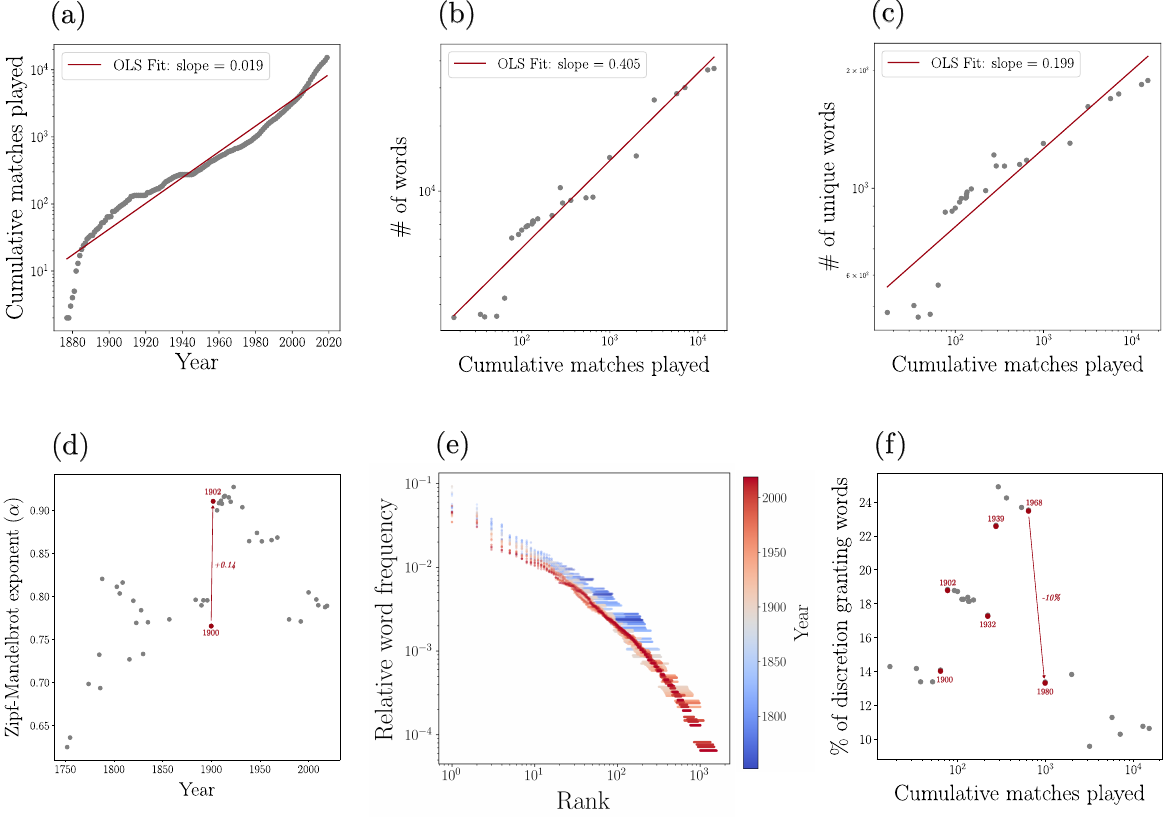}
    \caption{\textbf{Longitudinal text-based analyses of the Laws of Cricket.} (a) Cumulative matches played against year displaying an exponential in time, using match data from the ESPNCricInfo database across all men's formats of cricket \cite{espncricinfo_statsguru}, with $R^2 = 0.943$. Generally, the number of matches played is a more informative independent variable than time, and where possible we use this as the independent variable to assess aspects of rule change. This choice reflects the hypothesis that rule creation and revision respond to the outcomes and experiences generated by matches played under the existing rule set. (b) Number of words in the rule set against cumulative matches played, $R^2 = 0.935$. Over 268 years, the Laws of Cricket have had a 36-fold increase. (c) Number of unique words in the rule set over cumulative matches played, $R^2 = 0.837$. Notice that both number of words and number of unique words grow as a power law in the number of matches played. (d) Zipf-Mandelbrot exponent ($\alpha$) against year. (e) Rank-frequency distribution against year, normalized by the total number of words in the rule set. Earlier years are blue and more recent years are red.  (f) Percentage of discretion granting words over cumulative matches played \cite{vonWright1968-VONAEI}.}
    \label{fig:textual_analysis}
\end{figure}

We can use these text-based measures to answer a number of questions of interest. 
For example, a common criticism of bureaucracies is that they grow to become encumbered by excessive and restrictive rules. We are interested in investigating whether the cricket rule sets exhibit such a growth trajectory.
Whether the Laws of Cricket do suffer from regulatory bloat can be seen from how the rule books grow in time. We are also interested in whether cricket rule sets are becoming more systematized and formalized in time, as these are traits of bureaucratization \cite{weber2019economy,richerson2023institutional}. As we observe the trends over these various measures, we also can develop a sense of whether the changes to the cricket rule sets are gradual or sudden, and thus whether distinct phases and phase transitions exist across the dataset, as well as whether the rule sets evolve in the direction of granting discretion or constraining discretion over these distinct phases. 
Figure \ref{fig:textual_analysis} summarizes the main results of our textual analysis.

The number of words is an intuitive measure of the amount of regulation in a rule book---more words allowing for more regulations to be specified. 
The first observation in Fig.~\ref{fig:textual_analysis}(b) is a power-law relationship between the rule book size and the number of matches played with an exponent of 0.4---an instance of sublinear scaling, see Fig.~\ref{fig:textual_analysis}(a). Many subsequent figures use \textit{cumulative matches} as the independent variable; however match data was only available from 1877, so where the longitudinal trend of the preceding years (1752-1876) is important, we have kept time (year) as the independent variable. This indicates that as more matches are played, the number of additional words in the rule book per match decreases, but never saturates. Additionally, the number of unique words in a document is often seen as a measure of novelty \cite{loreto2016dynamics}. In Fig.~\ref{fig:textual_analysis}(c) we show that unique words grows as a power law in the number of matches played (recovering Heaps' law with a typical exponent of around 0.5). 



From the text statistics alone, we do not obtain a clear indication as to whether cricket rule sets become more systematized and formalized over time. On one hand, from Fig.~\ref{fig:textual_analysis}(e), we notice a great degree of visual similarity across the 268 years of normalized rank-frequency distributions.
This echoes recent findings of universality in rank-frequency structures of other complex systems \cite{holehouse2025generative}, driven by sublinear mechanisms of preferential attachment. This would seem to suggest that the functional form, and thus the lexical systematicity, of the rule sets remain relatively consistent over time. Figure~\ref{fig:textual_analysis}(d) shows the trend in the exponent parameter $\alpha$ of the Zipf-Mandelbrot fits \cite{mandelbrot1953informational} to the rank-frequency distributions of words in the rule sets over time (for the individual fits, please see Fig.~\ref{fig:all-zm-fits-p1} and \ref{fig:all-zm-fits-p2}). Across the entire corpus of cricket rule sets, the rank-frequency distribution is consistently well-described by a Zipf-Mandelbrot model, with $R^{2}$ values exceeding 0.96. On the other hand, in Fig.~\ref{fig:textual_analysis}(d) we show that the Zipf-Mandelbrot exponent fit longitudinally to the rule books is fairly inconsistent, varying between 0.7 to 0.9. This finding hints towards the punctuated and discontinuous nature of the evolution of the Laws of Cricket, revealed in the next section.

Even though the text-based measures are able to help us answer many important questions, they also have limitations. Importantly, there are structural properties to the rule sets that the text-based measures fail to capture. Cricket rules are organized in a hierarchical tree structure, with different categories of rules nested in other higher-level categories (see Fig.~\ref{fig:tree_nomenclature}). However, text-based measures are unable to capture these hierarchical relationships. Additionally, text-based measures are unable to capture the structural relationships \textit{between} rules. These structures form the backbone of each rule set, and are a necessary layer of abstraction for us to understand the evolution of the Laws of Cricket at a more fundamental level. In our next two sections, we will introduce a structural approach to analyzing the evolution of rule books, which reveals distinct phases in the Laws of Cricket.


\subsection*{Tree-based measures of evolution}



\noindent Whilst words in rule books contain the information required to play the game, there is no clear mapping from the semantics of the rule books to the complexity of the sport as practiced. A principled way to investigate the increasing complexity of sports rules, that we invoke here, involves extracting the hierarchies of rules contained within sports rule books, which reveals trees summarizing the structural complexity of the rules of the game. In a coarse way, one can think of the process of extracting rule hierarchies as quantifying the qualitative claims proposed by Daston in ref.~\cite{daston2022rules} on how \textit{thick} or \textit{thin} an individual rule is---the thinner and more conditional a rule is, the more nested and deep that rule is in the rule book. For example, in the 2019 rule book, Law 1.1 on the \textit{number of players}, which can by agreement be more or less than 11 players, is less specific and ``thicker'' than Law 19.2.2.2 on the placement of flags or posts at the boundaries of the field of play if the boundary is marked out by a white line.

\begin{figure}[ht]
    \includegraphics[width=.8\textwidth]{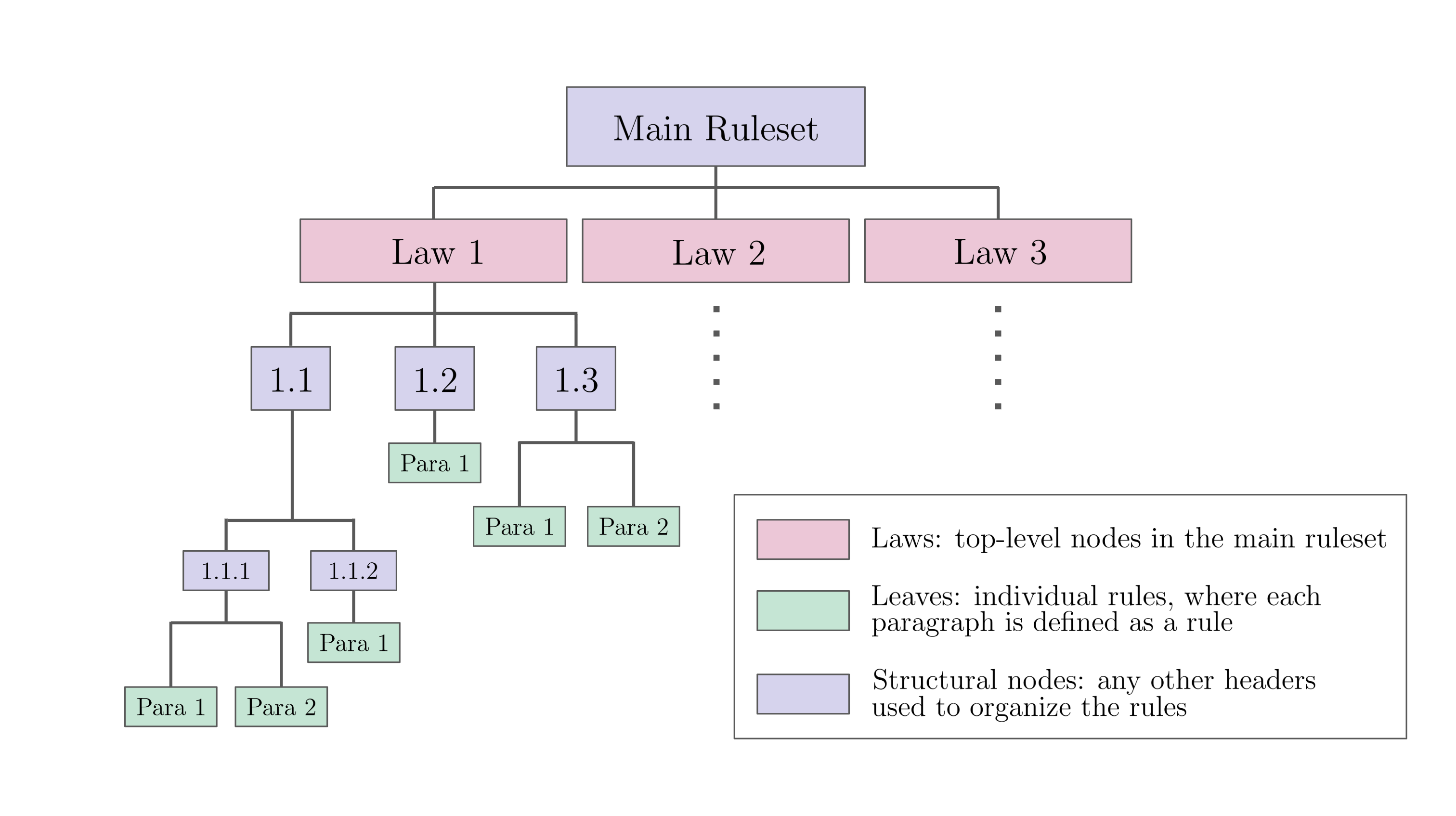}
    \caption{\textbf{Diagram of rule trees in terms of Laws (the highest-level nodes aside from the root node), rules (leaves), and structural nodes.} Also refer to the corresponding definitions in Box 1. The leaves, which we refer to as the rules, are determined by individual paragraphs, containing the semantics of the rules (also known as \textit{clauses} by rule makers). Each rule is a self-contained semantic packet defined by the rule makers.}
    \label{fig:tree_nomenclature}
\end{figure}

Each edition of the Laws of Cricket has a hierarchical structure, with enumerated clauses nested at multiple levels (see Fig.~\ref{fig:tree_nomenclature}). The placement of clauses within this hierarchy (and thus their relationships with respect to each other) is not arbitrary---it reflects an intentional choice of the rule makers on how to organize the semantic content of the rule book. For example, subclauses nested within the same parent clause typically address comparable issues and function as more specific elaborations of parts of the broader rule. 
However, there is no single necessary way to organize a set of rules. Thus, the structure of the rule set also reflects human judgment and institutional choices in deciding how to organize the rules. Therefore, studying the evolution of the tree structures gives us insight into \textit{both} the underlying relationships between rules, \textit{and} into how human institutions choose to organize complex bodies of regulations.  

\begin{figure}[ht]
    \includegraphics[width=1.0\textwidth]{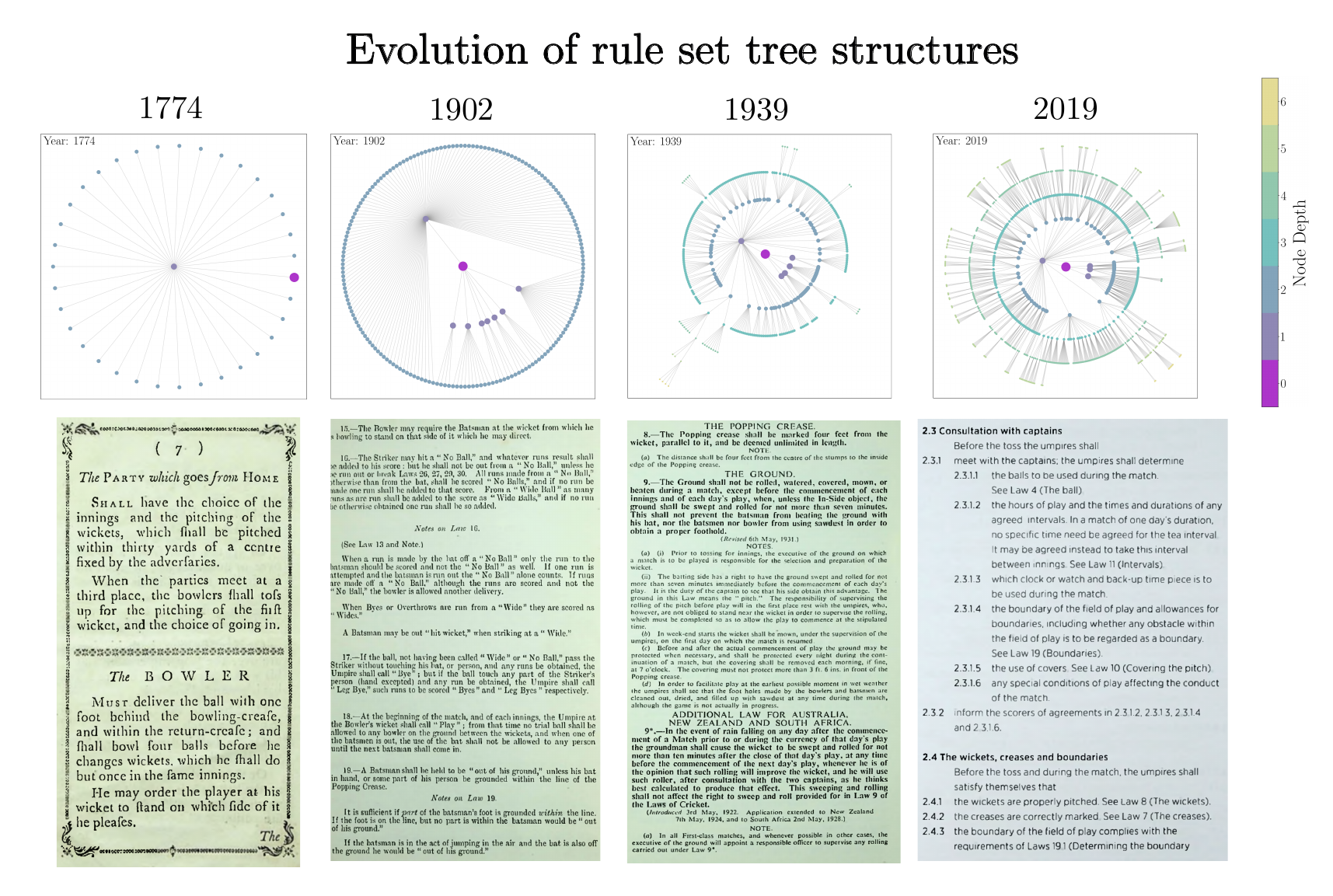}
    \caption{\textbf{Examples of the evolution of the rule trees from close to sport's inception (1774) to the modern era (2019).} Rule trees are shown side-by-side with samples from the rule books, exhibiting the relationship between the abstract rule trees and the underlying rule books. See Figures \ref{fig:all-years-ruleset-network-grid-p1} and \ref{fig:all-years-ruleset-network-grid-p2} for all the document trees in the dataset.}
    \label{fig:tree_evolution}
\end{figure}

At a high level, we want to know how these hierarchical structures grow in size. First, we want to understand the growth in the number of rules in time (i.e., the leaves of the tree). 
Next, we want to understand the growth in the number of nodes of the tree: which includes not only the rules themselves, but also the enumerations that act as organizing elements. 
In Fig.~\ref{fig:tree_evolution}, we show the correspondence between pages in the rule books and the resultant tree structures for a given year. These years have been chosen to emphasize key changes in rule book structures.

Apart from rule tree size, there are other properties of the tree that we are interested in. 
The depth of a tree is one of the measures we use. It refers to the maximum number of edges from the root node to any leaf in the tree. In our context of the cricket rule sets, an additional layer of depth is introduced when a clause is divided into multiple subclauses, typically to accommodate more specific elaborations that would be cumbersome to include within the original clause. Thus, we analyze the depth of the tree as a measure of the amount of \textit{specificity} in the rule set.

We also calculate the average branching factor of the trees. The branching factor of a node refers to the number of children that it has. For us, the branching factor of a node is the number of subclauses directly beneath it. We are interested in using the average branching factor of the tree (calculated for the non-leaf nodes of the tree) to measure the number of distinct elements that rule makers tend to group within a single category. This quantity may reflect both: (1) the inherent number of meaningful semantic distinctions within a given topic, and (2) cognitive limits on how many categories humans are comfortable grouping together before further subdivision becomes preferable. Examining the resulting branching factor lets us observe how these two forces interact in the organization of the rule set. 

\begin{figure}[ht]
    \includegraphics[width=0.8\textwidth]{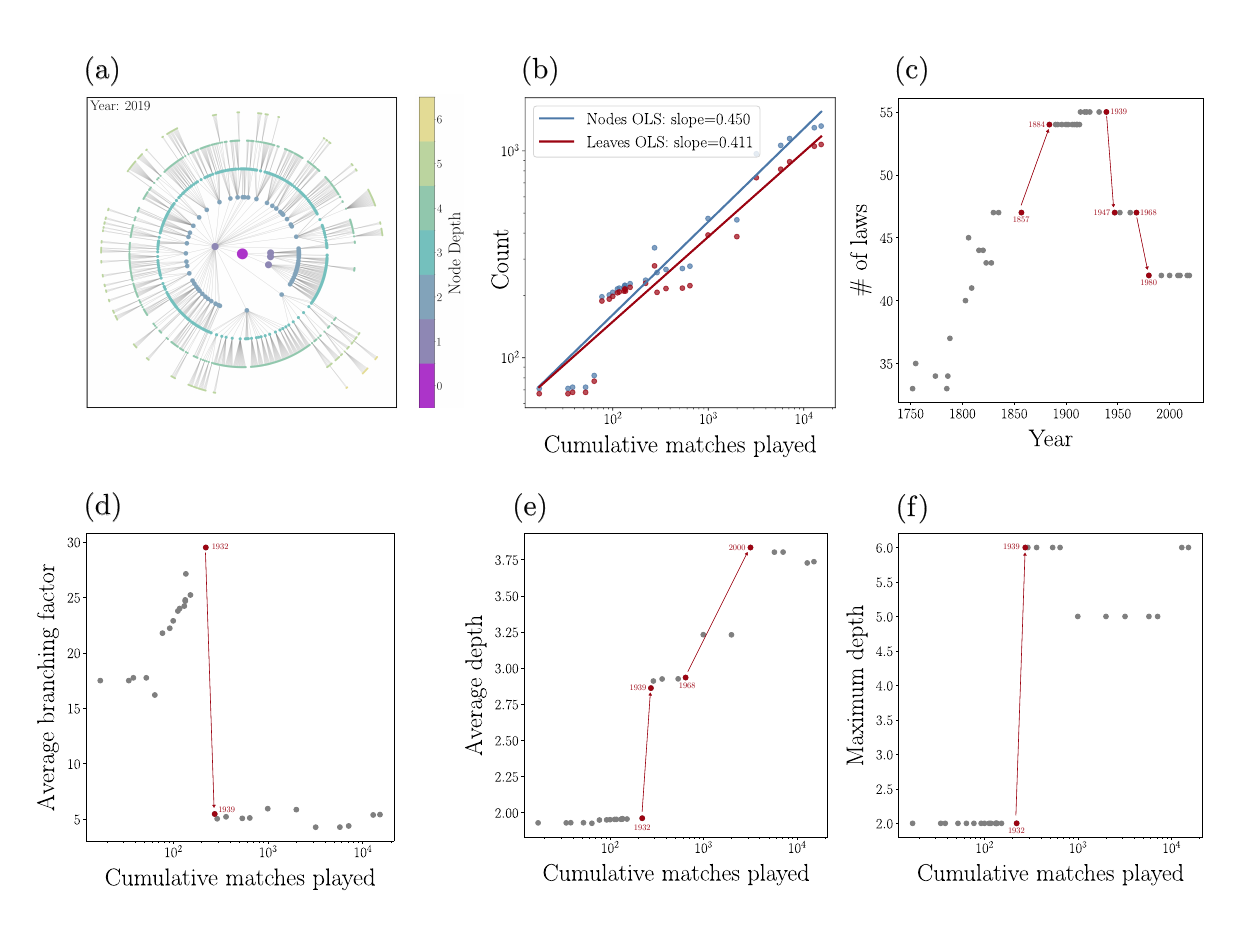}
    \caption{\textbf{Analyses of rule set tree structures.} (a) Radial visualization of the tree structure of the 2019 rule set, colored by node depth, using the \texttt{twopi} network layout. (b) Count of nodes and leaves over cumulative matches played; $R^2$ for number of nodes is 0.918 while $R^2$ for number of leaves is 0.890. (c) Number of Laws over year. (d) Average branching factor over cumulative matches played. (e) Average depth over cumulative matches played. (f) Maximum depth over cumulative matches played.
    }
    \label{fig:tree_analysis}
\end{figure}


In Figure \ref{fig:tree_analysis} we show the longitudinal analyses for the evolution of rule trees in the Laws of Cricket. Figure~\ref{fig:tree_analysis}(a) shows a radial representation of a rule tree. By 2019, there are many layers of depth in the cricket rule set that are all fairly well-developed. 
Fig.~\ref{fig:tree_analysis}(b) displays the trends of both the number of nodes and leaves with gameplay, and include power law fits to the data (which correspond closely with the growth of text size). Power-law fits between the number of leaves and nodes versus the number of matches played were found to have exponents of 0.41 ($R^2=0.890$) and 0.45 ($R^2=0.918$) respectively.

Figure~\ref{fig:tree_analysis}(c) displays the number of Laws---the number of top-level headers---over time. 
Given the exponential growth of the number of rules with time, it is remarkable that the number of Laws does not follow a monotonically increasing trend. The number of Laws steadily increases until 1857, after which it jumps to around 55 for around half a century. Thereafter, however, it decreases sharply from 1939 to 1947, and then again from 1968 to 1980. 
This indicates that standard models of tree and network growth, such as Yule processes \cite{aldous2001stochastic}, the Galton-Watson process \cite{harris1963theory}, and preferential attachment \cite{barabasi1999emergence,krapivsky2001organization}, cannot be used to understand the evolution of rule trees in the Laws of Cricket.

\begin{tcolorbox}[width=\textwidth,title={\textbf{Box 2: Diffusive interpretation of regulatory bloat in the Laws of Cricket}}]   
    \small{

    Assuming that the rate of new rule creation is proportional to the number of new situations encountered in the sport \cite{ostrom2014institutions}, we can consider diffusion in a \textit{space of situations}. A step in this space occurs every time a game is played, and the position of a diffusing particle represents the situation of a given game. The simplest case of a Brownian process describes a particle traversing an homogeneous $M$-dimensional space \cite{Haken1983Synergetics,redner2001guide}, which admits \smash{$r(t)\sim N(t)^{1/2}$}, where $r(t)$ is the typical radius explored by a particle in the space of situations. The typical radius explored is a measure of the novelty the system is expected to encounter as the process continues. Our empirical exponent (for leaves on trees) of 0.41 is less than 1/2, which means that the space of finding new situations in the game of cricket is a \textit{sub-diffusive} process \cite{metzler2000random}. Sub-diffusive dynamics can be seen through random walks on fractals wherein \smash{$r(t)\sim N(t)^{1/d_w}$}, where $d_w$ is the walk dimension. Therefore, the emergence of new rules in cricket can be thought of as the typical distance traversed on a fractal structure of dimension $d_w = 1/0.41 = 2.44$ \cite{barlow2006diffusions}. 
    
    $\quad$Under this narrative, the size of the rule book is not affected by the interactions between rules, only the number of games to have been played. 
    However, while $R(t)$ increases smoothly with $N(t)$, the changes in rule book structures contain discontinuities not visible at this level of the data. This can only be explained by considering complexity arising from having unintended interactions between the rules. We explore this point further in Box 3.
    }
\end{tcolorbox}

Figure~\ref{fig:tree_analysis}(d) plots the trend of average branching factor with gameplay. Figure~\ref{fig:tree_analysis}(e) and (f) show the trends of average depth and maximum depth of the tree with gameplay respectively. The average branching factor starts at approximately 15, and climbs up to 30 in the year 1932. However, the branching factor plummets to 5 in the year 1939, and has remained around that level ever since. This is an indication that the average number of subclauses in a clause decreases significantly from 1932-1939. This could be reflective of a decision of the rule makers to group subclauses in smaller groups. Even as the trees grow deeper from the year 1939 onward, the mean group size stays the same.

Turning our attention to tree depth, we find that the maximum depth of the rule set increases from 2 to 6 between the years 1932 and 1939. This means that by 1939, there is at least one section of the rule set that differentiates to 6 layers deep. The maximum depth then stagnates at the range of 5-6 thereafter. At face value, this might seem to suggest that after the significance increase in specificity and complexity of the cricket rules from 1932 to 1939, neither of them increased significantly thereafter.  We notice a different trend for the graph of average depth. Indeed, average depth does shoot up from 1932 to 1939, it only increases from 2.0 to 2.9, which reflects that a majority of branches in the cricket rules did not reach the maximum depth of 6. We then see the average depth of cricket rules increase in two bounds, all the way to 3.8 by the year 2000. This seems to suggest that the specificity and complexity of cricket rules continued to increase even after 1939. Indeed, when we inspect the radial tree visualizations over time (see Fig.~\ref{fig:tree_evolution}), we notice that most of the outer concentric circles in 1939 are sparsely populated, but get slowly filled in with successive years of cricket rule sets. Thus, we see specificity and complexity going from being an artifact of a single part of the cricket rules to being the norm across the cricket rules. 



\subsection*{Interdependency-network measure of evolution}


\noindent We now turn our attention to another important aspects of the Laws of Cricket—citations. Beyond acting as a means for organization, citations provide insight into patterns of interdependency and endogenous rule change that naturally arise within the game and its rules.
Rules connected through citations are more likely to interact or govern related aspects of the game, and are thus more likely to drive changes in one another. Rule changes affect gameplay, and these changes in gameplay in turn prompt adjustments to the rules \cite{march2000dynamics}.  Interdependency networks thus become a data-rich proxy for us to quantify the strength of these relations.

\begin{tcolorbox}[width=\textwidth,title={\textbf{Box 3: Interacting rules—phase transitions in the Laws of Cricket}}]   
    \small{
    While the growth of rule book size is well dictated by the number of matches played, the evolution of structure in the rule book is not smooth. As shown in Fig.~\ref{fig:tree_analysis}(d), and throughout the evolution in Fig.~\ref{fig:tree_evolution}, the rule hierarchies evolve with punctuated events delineating periods of smooth structural change. Prior to the phase transition in the average branching factor in 1939, the smooth change constituted a gradually growing branching factor. After the phase transition in 1939, the branching factor remains constant at around a value of $\sim5$. This discontinuous restructuring in the Laws of Cricket resembles a first-order dynamical phase transition \cite{canovi2014first,budich2016dynamical}---defined by the presence of discontinuities in an order parameter as a system evolves in time, which in our case is the branching factor of the rule tree. Additionally, prior to 1835, the Laws of Cricket did not contain citations between the Laws, whereas in the most recent datasets the Laws prolifically cite each other.

    $\quad$A side effect of decreasing the branching factor and increasing the depth of the rule trees is that rules become increasingly more specific. Newly evolved rules are more easily codified in these frameworks. In addition, existing rules are less likely to have unintended interactions with other rules due to more efficient codification. Concurrently, while rule trees become significantly deeper in time, and in a colloquial sense more modular, the interdependency networks become \textit{less modular in time}, shown in Fig.~\ref{fig:cit-ntwk-stat-summ}(e)-(f). This signifies different functions for the organization of the rule books through rule trees, and the interdependency of the rules themselves. While the interdependency networks show that the Laws of Cricket become more dependent on each other in time, with both lesser modularity and greater clustering, this interdependence is semantically separated by restructuring of the rules into more meaningful categories. Overall, this allows the game to become increasingly interconnected, while the rules maintain their legibility to laypeople.
    }
\end{tcolorbox}

\subsubsection*{Interdependency networks become increasingly complex in the Laws of Cricket}

\noindent We aggregate all citations at the level of Laws, from which we construct weighted, directed interdependency networks of each year's rule set. Due to earlier rule sets either lacking rule numbers or lacking regular expressions that are indicative of citations, we are only able to construct interdependency networks starting from 1835. Basic summary statistics for these networks can be found in Fig.~\ref{fig:cit-ntwk-suppl-fig1}. First, we want to estimate which functional forms best represent the in-weight and out-weight distributions of the interdependency network; this will allow us to construct model mechanisms on how these networks evolve. Past research has used in-degree distributions, out-degree distributions, and clustering coefficients to classify networks into networks of different types, e.g., scale-free networks, random graphs, and hierarchical networks \cite{Barabasi2004-bj}, and to construct narratives about regulatory evolution in gene-regulatory networks.

\begin{figure}[ht]
    \includegraphics[width=0.8\textwidth]{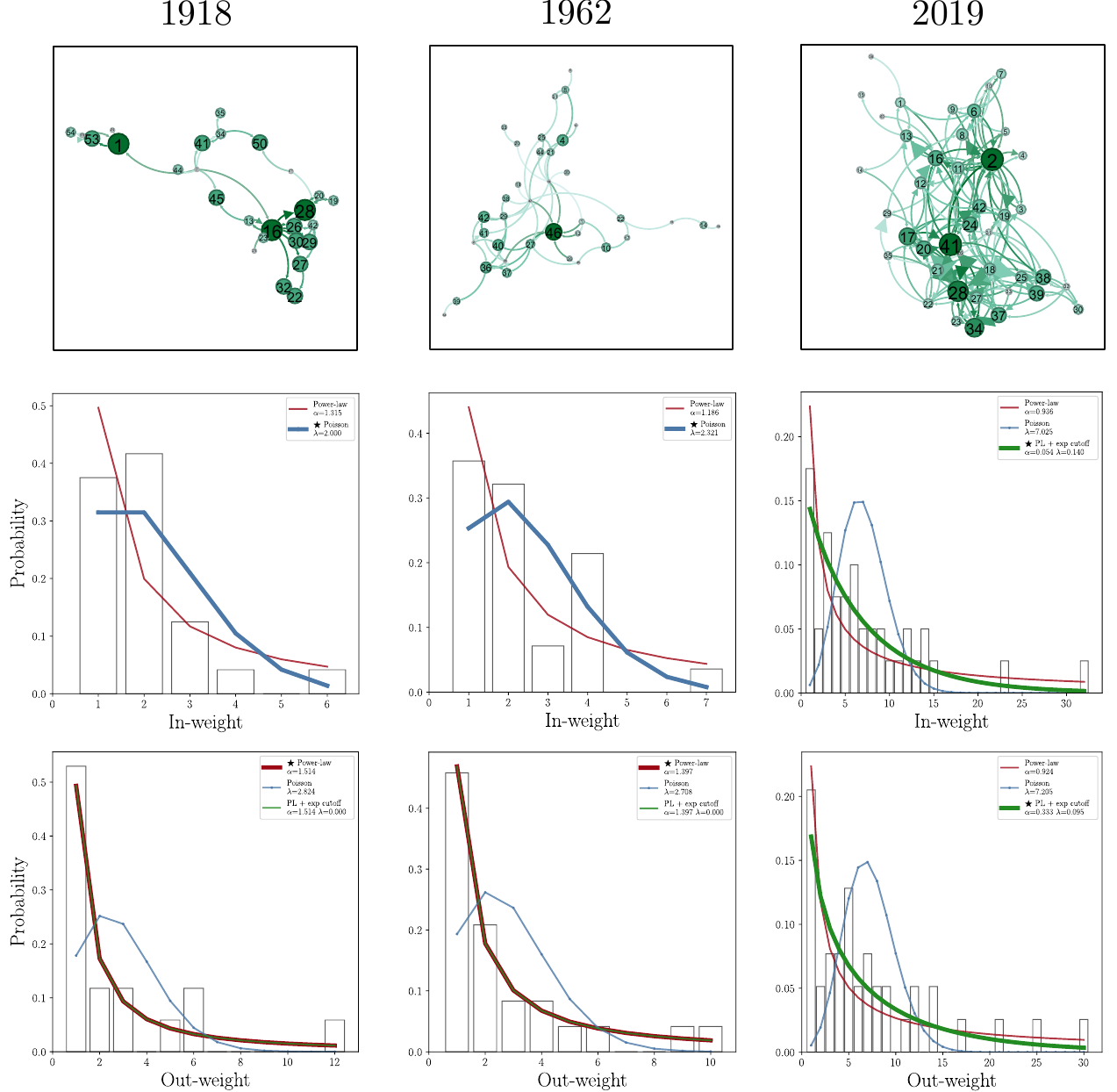}
    \caption{\textbf{Evolution of the interdependency networks from 1918 to 2019.} Row 1: interdependency network visualizations in Gephi for the years 1918, 1962, and 2019. Row 2: in-weight distributions for those years, with best fit distributions plotted over the empirical data. In the legend, the distribution labeled with a $\star$ indicates the maximum likelihood distribution chosen from model selection. The legend also indicates the best fit parameter values, whose distributions functional forms are specified in the \textit{Materials and Methods}. Row 3: out-weight distributions for those years.}
    \label{fig:citation_network_evolution}
\end{figure}

The first row in Fig.~\ref{fig:citation_network_evolution} illustrates three interdependency networks. The second and third rows in Fig.~\ref{fig:citation_network_evolution} contain the in-weight and out-weight distributions of the interdependency networks. The in-weight and out-weight of a node correspond to the total weight of incoming and outgoing edges respectively. Erdős–Rényi random graphs typically exhibit approximately Poisson degree distributions, whereas networks generated by preferential attachment mechanisms tend to exhibit power-law (scale-free) degree distributions. In practice, however, preferential-attachment networks are finite, and finite-size effects often cause their degree distributions to be better approximated by a power law with an exponential cutoff \cite{clauset2009power,tauber2014critical}. 
Thus, we fit Poisson, power-law distributions, and power-law distributions with exponential cutoffs to the empirically observed weight distributions for each year. Since the graph is directed, we look at both in-weight and out-weight distributions to see if they exhibit different behavior; using the Akaike information criterion (AIC) \cite{Akaike-info-crit} to identify the model that best fits the data.

\begin{figure}[ht]
    \includegraphics[width=1\textwidth]{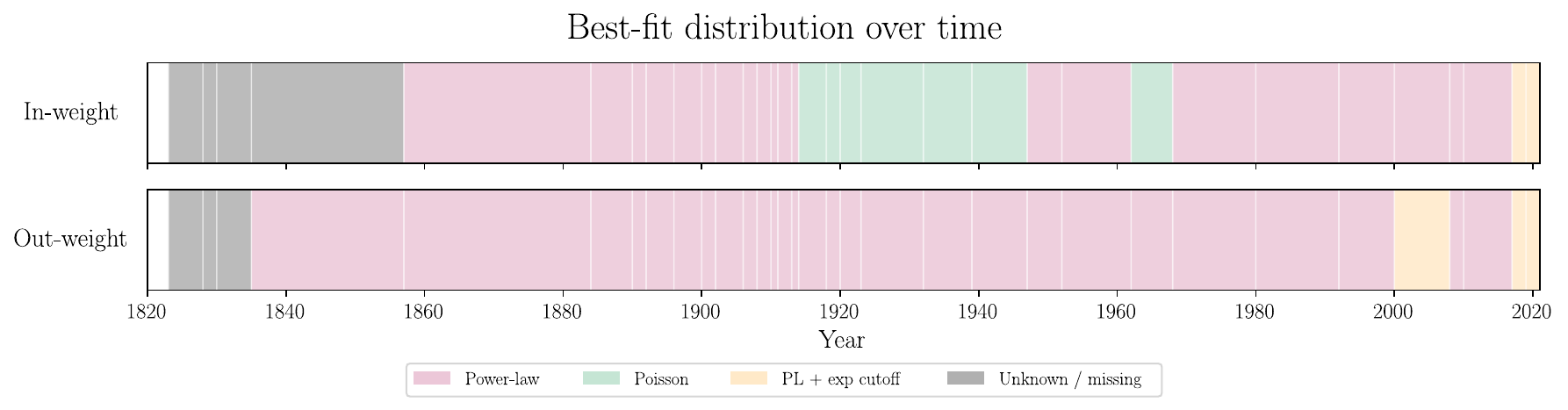}
    \caption{\textbf{Timeline indicating the best model fit for in-weight and out-weight over the duration of the rule set.}}
    \label{fig:best_fit_heatmap}
\end{figure}

Figure~\ref{fig:best_fit_heatmap} displays the best model fits to the in-weight and out-weight data across the duration of the rule sets. Overall, the power-law distribution is the best fit for the interdependency networks, for both in-weight and out-weight. However, two notable deviations emerge. First, for in-weight and out-weight, the power law with exponential cutoff outperforms the other two models for a few years in the last two decades of the rule set, even after the penalization from the AIC for having an extra parameter. Secondly, there are periods where in-weight (but not the out-weight) is best modeled by a Poissonian distribution; this indicates a difference during these periods in how edge tails (going out of citing Laws) and edge heads (going into cited Laws) are distributed among the Laws of Cricket. Nevertheless, the majority of the distributions are still best approximated by either a power-law distribution or a power-law distribution with an exponential cutoff. This is consistent with the family of scale-free networks, which commonly arise via the process of \textit{preferential attachment}, whereby highly connected nodes have an increased probability of acquiring additional links.

\begin{figure}[ht]
    \includegraphics[width=0.8\textwidth]{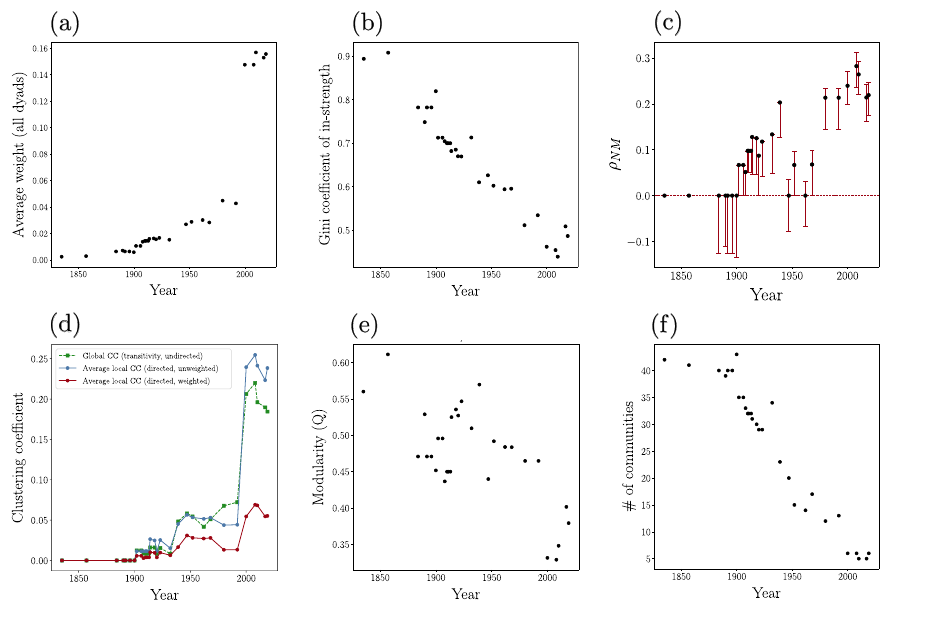}
    \caption{\textbf{Summary figure of interdependency networks statistics.} (a) Average edge weight, including dyads with zero-weight. (b)  Gini coefficient of in-strength and out-strength of the Laws over time. (c) Plot of the reciprocity of the weighted graph relative to a random baseline, $\rho_{NM}$, over time. (d) Various measures of clustering coefficient over time, aggregated at the network level. (e) Modularity over time. (f) Number of communities over time.}
    \label{fig:cit-ntwk-stat-summ}
\end{figure}

Having used the in-weight and out-weight distributions of individual interdependency networks to characterize the evolution of the interdependency network over time, we now turn to trends in classic network-level summary statistics to obtain a fuller picture of the evolution of law interdependence in cricket. In Fig.~\ref{fig:cit-ntwk-stat-summ} we present six plots of summary metrics as functions of time. First, in Fig.~\ref{fig:cit-ntwk-stat-summ}(a), we observe that the average weight of all dyads in the network increases steadily over the period leading up to 2000, after which it rises more sharply. This reflects the growing importance of rule interdependence over time. Second, in Figs.~\ref{fig:cit-ntwk-stat-summ}(b), we show that the Gini coefficients of the in-strengths of nodes systematically decrease over time, indicating that rule importance becomes more evenly distributed as the rule set evolves. A similar trend is also observed for out-strengths (not shown). Third, in Fig.~\ref{fig:cit-ntwk-stat-summ}(c), we examine how the reciprocity, $\rho_{NM}$, of the interdependency network—a measure of ``\textit{does the node I cite cite me?}'' introduced in ref.~\cite{Squartini2013-hi}—changes over time. We observe an overall increasing trend in $\rho_{NM}$, indicating that the interdependency network becomes more reciprocal over time. Notably, these measures of reciprocity are already normalized with respect to a random baseline with the same number of nodes and weighted edges. Thus, the observed increases are not simply a consequence of changes in network size or density, but instead reflect a genuine rise in reciprocal interactions between Laws.

Zooming out from the dyad level, we next use clustering coefficients to investigate closure at the local level. In Fig.~\ref{fig:cit-ntwk-stat-summ}(d), we examine how these coefficients evolve over time. We observe an increase in all measures of the clustering coefficient from 1932 to 1939, followed by a much steeper rise from 1992 to 2000. The weighted average local clustering coefficient exhibits less pronounced variation compared to the other two measures. As with the reciprocity observed at the dyadic level, we again find an overall increase in clustering over time. Taken together, these results indicate that the closure of the interdependency networks is increasing over time. 

At the network level, we quantify the separability of the network using modularity. Unlike clustering, modularity is not restricted to immediate neighbors—communities may include Laws that are not directly connected. We apply the Clauset–Newman–Moore greedy modularity maximization algorithm to identify communities in the network \cite{PhysRevE.70.066111}.  In Fig.~\ref{fig:cit-ntwk-stat-summ}(e), we observe an overall decreasing trend in modularity over time, and in Fig.~\ref{fig:cit-ntwk-stat-summ}(f) we observe that the number of communities detected by the algorithm decreases over time. Notably, this happens even during the period when the number of Laws is increasing, indicating that newly added Laws are increasingly absorbed into existing communities rather than forming new ones. In sum, this means that even as Laws are being agglomerated into communities, we still see an increase in the links between these communities. 

These findings present an apparent paradox: network closure increases over the duration of the rule set---as evidenced by rising reciprocity and clustering coefficients---while modularity simultaneously declines. However, the two trends are not contradictory, but complementary. Greater closure produces tighter local cycles and reinforcing loops that can drive endogenous rule change within clusters of closely related rules. Declining modularity, meanwhile, signals a proliferation of bridging edges that span community boundaries, ensuring that such change does not remain locally contained. Together, these structural properties create the conditions for rule change to both self-reinforce within clusters and propagate outward across the broader network.

\subsubsection*{Interdependency networks reveal changes in the cricket zeitgeist}

\noindent We now look at the importance of individual Laws using eigenvector centrality to quantify how the importance and interconnectedness of individual Laws evolves over time. For us, a Law's eigenvector centrality is determined by the importance of the Laws that cite it and how frequently they do so. By examining how the ranking of a Law based on its eigenvector centrality changes over time, we can make semantic claims about changes to its prominence as well.

\begin{figure}[ht]
    \includegraphics[width=1.0\textwidth]{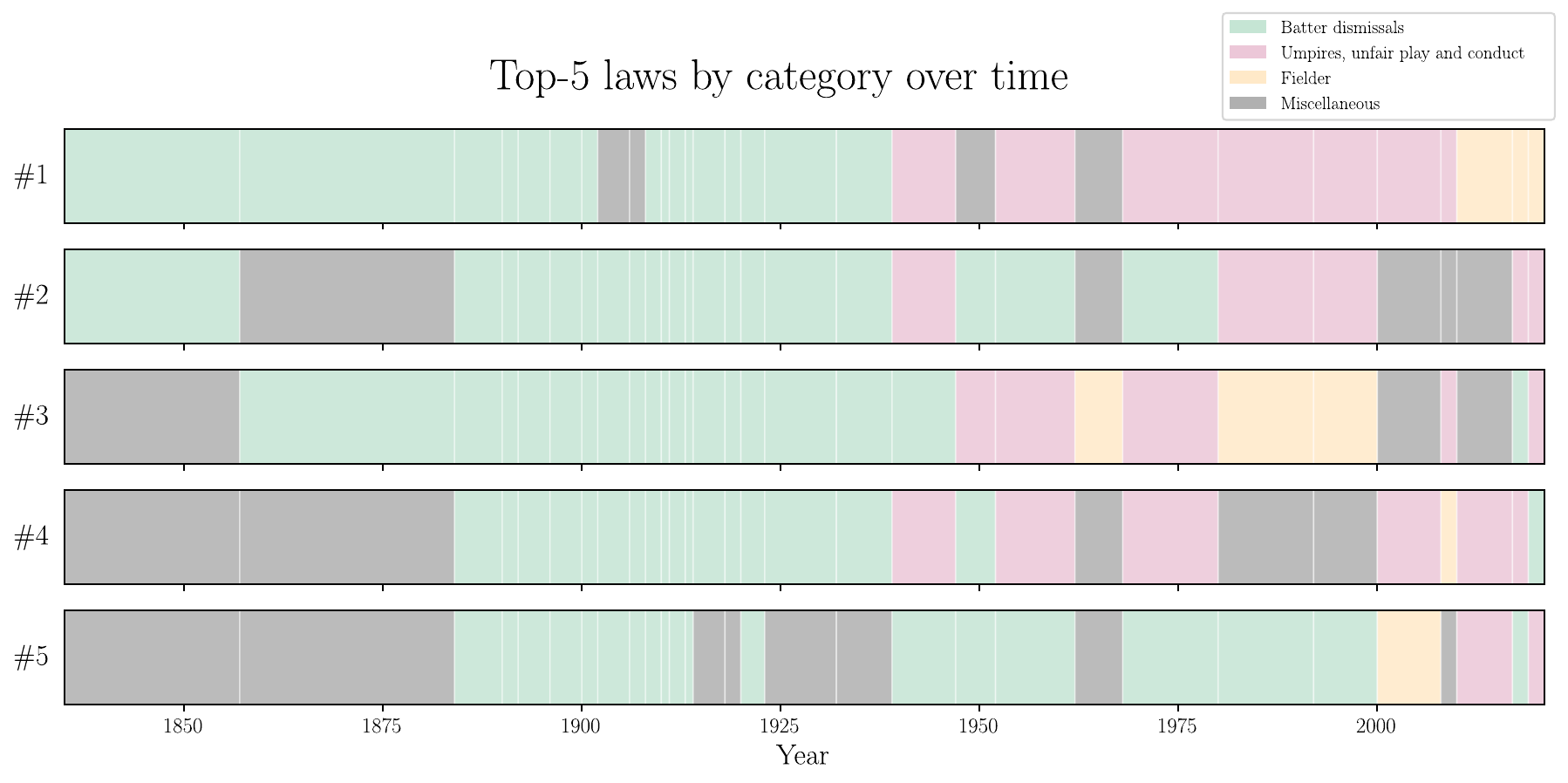}
    \caption{\textbf{Timeline of categories of laws in the top five, ranked by eigenvector centrality.} We see a systematic shift in importance of the laws, with earlier eras deeming Laws related to gameplay the most important, and more recent years deeming Laws related to officiating the game as more important.}
    \label{fig:top-five-ev-centrality}
\end{figure}

We calculate the eigenvector centrality distribution across nodes for each year of the Laws of Cricket, and then find the top five nodes per year. As in our case, previous research has conducted eigenvector centrality on weighted, directed networks \cite{Bonacich2007-kv,Jain2002-an,Taylor2017-xj}. For each year, we hand-label the top five Laws by comparing the Law number of the node back to its text label in the rule set PDF. For the sake of interpretability, we aggregate these Laws into four general categories: (1) batter dismissals, (2) umpires, unfair play and conduct,  (3) fielders, and (4) miscellaneous. These categories were determined by inspecting the Laws that appeared in the top five rankings. We visualize the evolution of the top five categories of rules in Fig.~\ref{fig:top-five-ev-centrality}. 

Over the course of evolution of the Laws of Cricket, we see a very clear shift in the importance of different classes of Laws. Prior to the middle of the 20th century, batter dismissals dominate the top-five Laws. However, from the latter half of the 20th century through to the present day, this shifts towards the Laws that govern umpires, unfair play and conduct, as well as those that govern the fielder. The shifting importance in the themes of individual Laws could be explained simply by the types of Laws that become difficult to interpret in different eras of the game. Early in the history of the Laws, core aspects of gameplay (e.g., batting) generated most interactions with other Laws. In the modern era, by contrast, interactions are increasingly driven by more nuanced aspects of the game (e.g., fair and unfair play). However, this could also be indicative of a shift in the relative importance of the various Laws of Cricket, and the increasing importance of regulating the regulators---a sentiment shared by Juvenal two millennia prior: \textit{``but who will watch the watchmen?''} \cite{juvenal_satires_vi}.

\section*{Discussion}

\noindent In this study, we have uncovered several core principles of regulatory evolution from the Laws of Cricket centered around the growth of rule book size, phase transitions in the rule set structures, and increased interdependence between rules. Importantly, these changes point to more general principles to be explored in regulatory systems writ large---principles that we illuminate in this discussion.

First, we showed over nearly three centuries that the Laws of Cricket grow exponentially in time and as a sub-linear power law in the number of matches played. This indicates that rule change happens primarily when people witness novel situations occurring in the game that are not anticipated by existing rules. Rule makers deliberate over these situations, and their deliberations sometimes lead to rule change. This appears to be a generic mechanism underlying regulatory evolution across diverse systems, including bureaucracies and genetic regulation. In the latter case there is, of course, no rule maker; instead, natural selection selects for useful regulatory mutations from large populations of cells. For cricket, as more matches are played the rate of discovery of novel situations decreases with time---an empirical law not observed for other rule systems such as the US Code which grows as $N^3$ with population size (see ref.~\cite{jeong2026dataset}) or the number of regulatory genes in prokaryotes and eukaryotes which grows as $N^{1.9}$ and $N^{1.3}$ with genome size respectively (see ref.~\cite[Fig.~1]{van2003scaling}). This indicates that while the mechanism of generating new rules based on encountered novelties may be universal, the details dictating how fast rules grow are highly system specific.

Second, we observe a phase transition in the organization of the Laws of Cricket. Prior to 1939, the branching factor of the rule trees increased rapidly, after which it stabilized around a value of (around) five. This increase in rule depth (and subsequently rule specificity) resembles other systems of regulation, and provides some context to qualitative claims made by Daston in ref.~\cite{daston2022rules} regarding the thickness of rules. For example, the prokaryotic-to-eukaryotic transition in biology was accompanied by a qualitatively observed increase in regulatory depth: more enhancers across the genome leading to more factors that can activate or suppress individual genes; chromatin actively restricts the accessibility of regions of the genome; insulator proteins (e.g., CTCFs) isolate regions of the genome preventing interactions outside that region. In short, regulation in eukaryotes is much more specific than in prokaryotes, similar to what we see in the evolution of the Laws of Cricket. With reference to Daston, our quantitative analyses show that not only do thick (broad) and thin (specific) rules exist within the same rule set, but that in general there is a trend for thick rules to be replaced by a multitude of thin rules. Broad rules are repeatedly confronted by exceptions, which become codified as an increased number of fine-grained and specific rules. In general, our analyses provide evidence that increasingly conditional and thinner rules are the natural consequence of the regulatory evolutionary process.

Third, we found that citations in the rule books paint a complex picture of how interdependencies form as rule books evolve. While later rule books have highly clustered interdependency networks, the modularity of those networks also decreases. This indicates that as the Laws evolve they become increasingly interconnected and themes in the Laws are harder to disentangle. We observe similarities to gene-regulatory networks in that the in-weight and out-weight distributions between Laws are generally power-law \cite{Barabasi2004-bj}, suggestive of preferential-attachment-like mechanisms being responsible for the emergence of regulatory structure in both cases. Importantly, the interdependency networks also allowed us to see the changing importance of themes in the rule book. Early on, rule books were dominated by rules related to aspects of gameplay, but later it is found that the most vital rules center around the regulation of the game itself---Laws related to officiating and unfair conduct. This finding, in the context of governmental bureaucracies, makes sense---interactions between rules and non-trivialities in their implementation lead to the proliferation of ``rules about rules''. In sum, cross-system comparisons suggest a common pattern of increasing regulatory interdependence, greater clustering, and shifting functional priorities as regulatory systems evolve.

Regulatory systems evolve from simple lists of broadly interpretable rules into increasingly conditional, interconnected, and deeply structured regulatory architectures.
There is fertile ground for our analyses to be replicated and extended to other longitudinal rule sets, across many other domains in the social and biological sciences. Our study provides a baseline dataset and set of analyses that can be extended across other rule domains, hopefully providing further evidence for the principles of regulatory evolution.

\section*{Materials and Methods}

\subsection*{The Laws of Cricket Data Set}



\begin{figure}[ht]
    \includegraphics[width=1.0\textwidth]{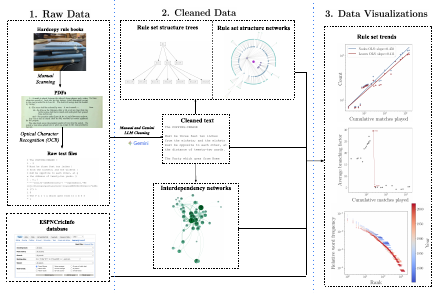}
    \caption{
    \textbf{Flow chart depicting the steps in the data pipeline.} We cleaned the raw text data and derived two alternate representations of the cleaned text data: (i) the rule set structure trees/networks, and (ii) the interdependency networks. Finally, we visualized the trends for various metrics of these different representations of the cricket rule set.}
    \label{fig:flow-chart}
\end{figure}

\noindent We collected all forty-four editions of the Laws of Cricket, spanning 1752-2019 (268 years), from the library at the Marylebone Cricket Club at Lord's Cricket Ground in London. All rule books that were available from the archives were scanned with a CZUR ET18 Pro Professional Document Scanner, with built-in auto-flattening software. These editions were scanned at the premises (over a three day period in September 2024), where optical-character recognition (OCR) software was then used to convert the scanned documents into machine-readable text. We cleaned this text with Google's Gemini AI. From the AI-cleaned texts, we then extracted the word statistics, hierarchical relationships between the rules (tree structures), and interdependencies between the rules as measured by the interdependency networks between Laws in the rule books. A summary of this process from data collection to construction of the processed data is given in Fig.~\ref{fig:flow-chart}. For more details, please refer to the Materials and Methods. For access to the data in raw, text-corrected, and processed forms, please see the\textit{ Data Availability Statement} below at the end of the text.

\subsection*{Processing of the Raw Cricket Rule Books}


\noindent Having all the scanned PDF files of the raw cricket rule books, the first step was to decide which version of the yearbook to keep when multiple versions existed for the same year. There were different publications that we found—some containing rules for a single year, some containing rules for multiple years—and we had to determine which version to use as the data point for that year in our analysis. We thus developed heuristics to make this determination, such as whether the rule set was a primary source (as opposed to part of a compilation), as well as whether its contents and structure made sense chronologically given the rule sets that came before and after it.

The next step was to decide which sections of the rule set to keep for our analysis. Over these centuries, these rule sets often included some additional segments: ranging from segments on the dimensions of the balls or wickets, to segments on Laws for specific countries, to segments on rules of county cricket that centered more on how players can qualify for certain competitions. We developed heuristics to determine which segments to keep, such as prioritizing rules related to gameplay, as well as factoring in whether the rules were in the main enumerated rule set or in additional segments at the end. See the Data Appendix (data\_appendix.xlsx) in the GitHub repository for a detailed description of the aforementioned heuristics.

After finalizing the rule set version for each year and the contents to keep for each of them, we did a manual check of the PDF files to check for strange anomalies (e.g., missing pages, pages wrongly ordered, errors in the Optical Character Recognition output). In most cases, this just involved deleting unnecessary segments; in rare instances where there was missing content, we had to look at the trends of the corresponding segments in the preceding and following rule sets to interpolate the missing content contained, and imputed it into the text file.

The final step was to use the Gemini API to do the remaining cleaning, with the following prompt: ``Please correct the following OCR-extracted text by fixing typos, spacing, and punctuation. Ensure that the cleaned text remains true to the original meaning without adding or removing content. Return only the cleaned text without additional comments. Original text: {text}''.\newline
This gave us machine-readable~\texttt{.txt} files with the text for each of the years with cricket rule sets, that we would carry on to use for constructing the rule book trees and interdependency networks. A summary of this processing procedure is provided in Fig.~\ref{fig:flow-chart}.

\subsection*{Fitting Zipfian Distributions to the Rule Set Data}
\noindent We first used the \texttt{nltk} Python package to clean and tokenize the text of each rule set. Next, we obtained the normalized rank-frequency distribution of the words in each rule set, using the relative frequencies of words rather than raw counts. We then fit a Zipf-Mandelbrot curve $f(r) \propto (r+\beta)^{-\alpha}$ to this rank-frequency distribution \cite{piantadosi2014zipf}, and obtained the parameters $\alpha$ and $\beta$ along with $R^2$ goodness-of-fit. We plot the exponent $\alpha$ over time in Fig.~\ref{fig:textual_analysis}(d), and we show all Zipf-Mandelbrot fits in Figs.~\ref{fig:all-zm-fits-p1} and \ref{fig:all-zm-fits-p2}.

\subsection*{Making the Rule Book Trees from Document Structure}
\noindent Different years of the rule books had different numbering schemes to codify their document structure. For example, the 1980 rule set was enumerated via a combination of numbers, letters and roman numerals to indicate different levels of depth in the rule set (e.g. ``8(a)(i)''), while the 2019 rule set was enumerated via a decimal system (e.g. ``19.1.1.7''). In order to extract their rule set structure trees, we needed algorithms that used regular expression matching to identify the aforementioned expressions signifying enumeration. However, these differences in numbering schemes meant that different algorithms were required for each. 

Fortunately, rule books from the same time period often shared a numbering scheme. We wrote different algorithms for different time periods of rule books grouped by the numbering schemes they used, and automatedly extracted the rule set structure for the whole dataset. This structure was saved in the~\texttt{.yaml} file format, a data format that allows for easy storage of hierarchical data. 

After that, we used the \texttt{anytree} Python package to convert the~\texttt{.yaml} file data into a~\texttt{.html} visualization of the tree structure. As a correctness check, for each of the~\texttt{.html} visualizations, we did a comparison with the original PDF of the rule set to check if there were discrepancies. For examples, please see Fig.~\ref{fig:tree_evolution}. This was first used to identify and rectify issues with the algorithms used to extract the rule set structures from the rule set text files. Often, however, there were minor errors in the cleaned text files themselves (generated by Optical Character Recognition and processed through Gemini) that distorted the expressions used to identify enumeration and made the algorithm unable to detect them. In these instances, we rectified the issues in the text files themselves before re-generating the tree.

Although visualizing the rule set structures as trees was intuitive, one limitation was that the trees were extremely long horizontally, and hence it only the tree visualization was only helpful when zooming in on a particular segment of the tree. In other words, it was difficult to gain any insight from looking at the tree as a whole. To address this, we generated a network visualization of each tree in a radial structure: generating visualizations separately with the \texttt{kamada-kawai} layout and the \texttt{twopi} layout from \texttt{NetworkX}. 

Lastly, one degree of freedom we introduced was a distinction between ``original'' and ``flattened'' rule set structure trees. 
We were not interested in merely how deep the tree is purely in terms of syntax: instead, we wanted to use the syntactic depth of the tree as a proxy for the degree of specificity of rules. In our original tree configuration, enumerating the paragraphs with numbers added an extra layer of depth to the tree, causing a jump in average depth at around 1828. However, merely tacking a number onto each paragraph was extremely unlikely to correspond to an increase in specificity of the rules. Hence, we decided to create a ``flattened'' version of the rule set, that places rule paragraphs on the same level of depth of the tree as the enumerations they correspond to, rather than having an extra level of depth for the paragraphs. This would ensure that our measure of the tree depth syntax-wise would actually correspond to the quantity we were interested in measuring, the degree of specificity of the rules. 

\subsection*{Making the Interdependency Networks}


\noindent Again, different years of the Laws of Cricket recorded citations in different ways. Initial attempts to use Generative AI to detect citations were unsuccessful, hence we resorted to using regular expression matching.

First, we adapted the code from parsing the document structures, and used it to match every rule in the rule set with the rule text that it contained. Next, we used regular expression matching to identify citations in each block of rule text, recording both the citing rule and the rule being cited. We then aggregated these citations at the Law level, recording the total citations from rules in one law to rules in another law. After that, we constructed a directed, weighted graph in \texttt{NetworkX} where the nodes represented the Laws, and the weights of the directed edges were the total citations from the Law represented by the source node to the Law represented by the target node. Lastly, we used network visualization software Gephi to visualize the network, determining the size and color of the nodes based on their degree, for examples please see Fig.~\ref{fig:citation_network_evolution}(top row). 

\subsubsection*{Model-Fitting to In- and Out-Weight Distributions}

\noindent After generating and visualizing the interdependency networks, we moved on to our analysis of the networks. We obtained the in-degree, out-degree, in-weight and out-weight distributions for each year's interdependency network. Then, we fit Poisson, power-law, and power-law with exponential cutoff distributions to the observed data, which have the respective functional forms:
\begin{align}
    P_{1}(x) &= \lambda^x e^{-\lambda}/x!,\\
    P_{2}(x) &\sim x^{-\alpha},\\
    P_{3}(x) &\sim e^{-\lambda x} x^{-\alpha},
\end{align}
where the latter two distributions are specified up to their normalization constants. We used the Akaike-information criterion (AIC) to choose the best fit between the proposed functional form and the data. This ensures that the two-parameter power-law distribution with exponential cutoff does not get an unfair advantage over the single-parameter power-law and Poisson distributions.

It is worth noting that we fit all our models on strictly positive support values---nodes with degree/weight equal to 0 were excluded from consideration. The reason for this was twofold: (1) power-law distributions are undefined at $x=0$, which makes it impossible to include nodes with degree/weight equal to 0 in the fitting process, and (2) we can reasonably consider nodes with 0 degree or weight to not be a part of the interdependency network, since they are neither cited nor contain citations. For consistency, we wanted to make sure all distributions were fit on the same support, hence these nodes were removed for all three distributional fits. Additionally, we normalized distributional fits over their observed support, rather than their full theoretical domain (i.e., the positive integers). This ensures that all three distributions were evaluated as conditional distributions over the same support, making their AIC scores directly comparable. 

\subsubsection*{Eigenvector Centrality}

\noindent For the eigenvector centrality, we computed eigenvector centrality both with and without edge weights factored in. The rankings of the Laws (top-five and individual case studies) were constructed based on eigenvector centrality with edge weight factored in. There were a couple of complications for the labeling of Laws with their eigenvector centrality. The first was that the Law number of a Law changes over time. For example, the Law ``Caught'' is Law 22 in 1896, Law 35 in 1947, Law 32 in 1980 and Law 33 in 2019. Additionally, not all Laws were stable over time. For example, over the lifespan of the Laws of Cricket, ``The Fieldsman'' was renamed to ``The Fielder'', ``Handled the ball'' was subsumed into ``Obstructing the field'', ``Appointment of Umpires'' was subsumed into ``The Umpires''. These factors combined meant that there was no clean automated way to track where a single Law was in the rule sets over time. Hence, all labeling had to be done manually, by looking through the rule set PDFs and identifying the rule of interest. 

We constructed the table of the top five ranked Laws over time. For each year, we took the eigenvector centrality of the nodes in the interdependency network and ranked them in decreasing order. We then took the Law numbers of the top five Laws and searched them up in the rule set PDFs to identify the name of each Law. We did this for all years of the rule sets that we had citation data for. While each of the Laws in later years had a clear title, there were instances in earlier rule sets where either (1) the title would contain multiple Laws, or (2) the Law had no title at all. This made it hard to assign a precise label to the Law---however this was still enough for us to classify the Law into one of the four categories that we developed.  



\subsubsection*{Some Concrete Examples of Eigenvector Centrality}
\begin{figure}[ht]
    \includegraphics[width=.6\textwidth]{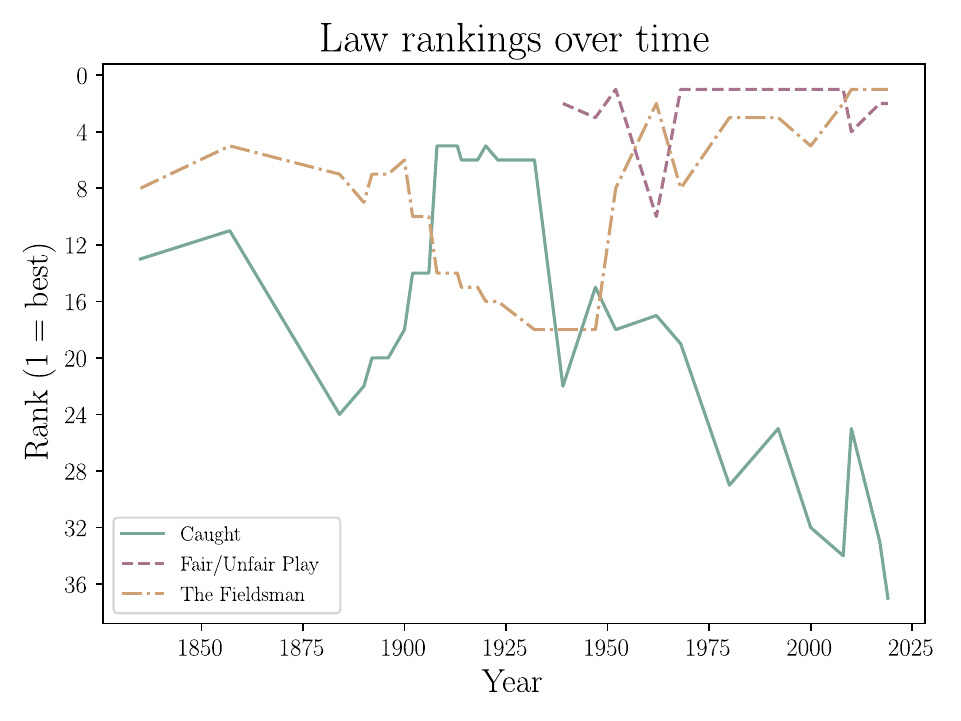}
    \caption{\textbf{Ranking of ``Caught'', ``Fair/Unfair Play'' and ``The Fieldsman'' over time.} Respectively, these show examples of: Laws that start out prominent but become less relevant in recent times; laws that remain important since their introduction; and laws that fluctuate in their importance over time.}
    \label{fig:three-case-studies}
\end{figure}

\noindent For concreteness, in Figure \ref{fig:three-case-studies} we chose three specific Laws as case studies and tracked their rankings over time. Each Law exemplifies the aforementioned trends in the Law category that it belongs to. The Law ``Caught'' regulates the most common method of batter dismissal, where the ball is caught after being hit before it touches the ground. While it experiences a period of brief popularity from 1900 to 1923, we see a marked and steady decline in its ranking from 1923 onward. The Law ``Fair/Unfair Play'' contains rules outlining conventions of the game that protect the ``Spirit of Cricket''. While this Spirit of Cricket lacks a precise definition, it is explained in the Preamble to the Laws of Cricket \cite{Preamble}, and contains provisions on issues such as ball tampering and distracting the opposition. The Law was only introduced in the later years of the Laws of Cricket, from 1939 onwards---but ever since its introduction, was consistently one of the highest-ranked Laws. The Law ``The Fieldsman'' contains rules regarding the fieldsmen, who are the players that try to collect the ball after it is struck by the striking batter, so as to limit the number of runs scored by the striker or to get a batter out. This Law ranked fairly highly in the beginning (6th-8th), before dipping in importance from 1902 to 1947. However, it regained ranking thereafter, eventually becoming the most important Law in the game. 

\section*{Acknowledgments}
\noindent JH would like to thank the Marylebone Cricket Club, in particular Alan Rees, for their hospitality and expertise in the cricket rule books whilst JH conducted the data collection. JH would also like to thank Marina Dubova and Tom McCarthy for their curiosity, ideas, and motivation.
DC would like to thank the Santa Fe Institute education programs, funded by the Darla Moore Foundation and the McKinnon Family Foundation. DC and JH would like to thank Jeremy Van Cleve, Simon DeDeo, and Harrison Hartle for useful discussions. The authors would like to acknowledge the support of the National Science Foundation Grant Award Number EF--2133863. HY ~acknowledges the NRF Global Humanities and Social Sciences Convergence Research Program (2024S1A5C3A02042671) and the support from the Institute of Management Research at Seoul National University.

\section*{Author Contributions}
\noindent Conceptualization: JH, with input from all authors. Preliminary computational analysis: JH. Extensive computational analysis: DC. Data acquisition: JH. Data processing: JH and DJ. Supervision: JH, with input from GBW and HY. Consultation on data processing pipeline: DJ. Funding acquisition: HY, GBW and CK. Figures: DC. Contributed computational analytic tools: JH and DC. Writing--original draft: JH and DC. Writing--review and editing: all authors.

\section*{Data Availability Statement}
\noindent The data, figures, and code used to generate the data and figures are all available at GitHub: \href{https://github.com/ZaniuSchiaaa/cricket-rule-evolution}{\textcolor{blue}{https://github.com/ZaniuSchiaaa/cricket-rule-evolution}}.

\bibliographystyle{unsrt}
\bibliography{biblio}

@book{heinrich2012regulation,
  title={The regulation of cellular systems},
  author={Heinrich, Reinhart and Schuster, Stefan},
  year={2012},
  publisher={Springer Science \& Business Media}
}

@article{yang2024regulatory,
  title={Regulatory Functions from Cells to Society},
  author={Yang, Vicky Chuqiao and Kempes, Christopher P and Redner, S and West, Geoffrey B and Youn, Hyejin},
  journal={arXiv preprint arXiv:2409.02884},
  year={2024}
}

@article{schultz2026topological,
  title={Topological Approaches in Animal Comparative Genomics},
  author={Schultz, Darrin T and Simakov, Oleg},
  journal={Annual Review of Animal Biosciences},
  volume={14},
  number={1},
  pages={17--48},
  year={2026},
  publisher={Annual Reviews}
}

@book{daston2022rules,
  author    = {Daston, Lorraine},
  title     = {Rules: A Short History of What We Live By},
  series    = {The Lawrence Stone Lectures},
  publisher = {Princeton University Press},
  address   = {Princeton, NJ},
  year      = {2022},
  isbn      = {9780691254081},
  pages     = {384},
}

@book{march2000dynamics,
  author    = {March, James G. and Schulz, Martin and Zhou, Xueguang},
  title     = {The Dynamics of Rules},
  publisher = {Stanford University Press},
  address   = {Stanford, CA},
  year      = {2000},
  isbn      = {9780804736754}
}

@article{clauset2009power,
  title={Power-law distributions in empirical data},
  author={Clauset, Aaron and Shalizi, Cosma Rohilla and Newman, Mark EJ},
  journal={SIAM Review},
  volume={51},
  number={4},
  pages={661--703},
  year={2009},
  publisher={SIAM}
}

@book{tauber2014critical,
  title={Critical Dynamics: a Field Theory Approach to Equilibrium and Non-equilibrium Scaling Behavior},
  author={T{\"a}uber, Uwe C},
  year={2014},
  publisher={Cambridge University Press}
}

@article{canovi2014first,
  title={First-order dynamical phase transitions},
  author={Canovi, Elena and Werner, Philipp and Eckstein, Martin},
  journal={Physical Review Letters},
  volume={113},
  number={26},
  pages={265702},
  year={2014},
  publisher={APS}
}

@article{budich2016dynamical,
  title={Dynamical topological order parameters far from equilibrium},
  author={Budich, Jan Carl and Heyl, Markus},
  journal={Physical Review B},
  volume={93},
  number={8},
  pages={085416},
  year={2016},
  publisher={APS}
}

@incollection{barlow2006diffusions,
  title={Diffusions on fractals},
  author={Barlow, Martin T},
  booktitle={Lectures on Probability Theory and Statistics: Ecole d'Et{\'e} de Probabilit{\'e}s de Saint-Flour XXV—1995},
  pages={1--121},
  year={2006},
  publisher={Springer}
}

@article{barabasi1999emergence,
  title={{Emergence of Scaling in Random Networks}},
  author={Barab{\'a}si, Albert-L{\'a}szl{\'o} and Albert, R{\'e}ka},
  journal={Science},
  volume={286},
  number={5439},
  pages={509--512},
  year={1999},
  publisher={American Association for the Advancement of Science}
}

@article{metzler2000random,
  title={The random walk's guide to anomalous diffusion: a fractional dynamics approach},
  author={Metzler, Ralf and Klafter, Joseph},
  journal={Physics Reports},
  volume={339},
  number={1},
  pages={1--77},
  year={2000},
  publisher={Elsevier}
}

@book{Haken1983Synergetics,
  author    = {Hermann Haken},
  title     = {Synergetics: An Introduction: Nonequilibrium Phase Transitions and Self-Organization in Physics, Chemistry and Biology},
  series    = {Springer Series in Synergetics},
  publisher = {Springer-Verlag},
  address   = {Berlin, Heidelberg},
  year      = {1983},
  edition   = {Revised, enlarged ed.},
  isbn      = {978-0412721601}
}

@book{redner2001guide,
  title={A guide to first-passage processes},
  author={Redner, Sidney},
  year={2001},
  publisher={Cambridge University Press}
}

@article{ostrom2014institutions,
  title={Do institutions for collective action evolve?},
  author={Ostrom, Elinor},
  journal={Journal of Bioeconomics},
  volume={16},
  number={1},
  pages={3--30},
  year={2014},
  publisher={Springer}
}

@article{krapivsky2001organization,
  title={Organization of growing random networks},
  author={Krapivsky, Paul L and Redner, Sidney},
  journal={Physical Review E},
  volume={63},
  number={6},
  pages={066123},
  year={2001},
  publisher={APS}
}

@book{harris1963theory,
  title={The theory of branching processes},
  author={Harris, Theodore Edward and others},
  volume={6},
  year={1963},
  publisher={Springer Berlin}
}

@article{loreto2016dynamics,
  title={Dynamics on expanding spaces: modeling the emergence of novelties},
  author={Loreto, Vittorio and Servedio, Vito DP and Strogatz, Steven H and Tria, Francesca},
  journal={{Creativity and Universality in Language}},
  pages={59--83},
  year={2016},
  publisher={Springer}
}

@article{holehouse2025generative,
  title={A generative model of function growth explains hidden self-similarities across biological and social systems},
  author={Holehouse, James and Redner, S and Yang, Vicky Chuqiao and Krapivsky, PL and Arroyo, Jose Ignacio and West, Geoffrey B and Kempes, Chris and Youn, Hyejin},
  journal={arXiv preprint arXiv:2509.14468},
  year={2025}
}

@misc{espncricinfo_statsguru,
  title        = {ESPNcricinfo Statsguru: Cricket Statistics Database},
  author       = {{ESPNcricinfo}},
  year         = {2026},
  howpublished = {\url{https://stats.espncricinfo.com/ci/engine/stats/index.html}},
  note         = {Accessed: 2026-03-09}
}

@book{richerson2023institutional,
  editor    = {Richerson, Peter J. and Bednar, Jenna and Currie, Thomas E. and Gavrilets, Sergey and Wallis, John Joseph},
  title     = {Institutional Dynamics and Organizational Complexity: How Social Rules Have Shaped the Evolution of Human Societies Throughout Human History},
  year      = {2023},
  publisher = {Cultural Evolution Society},
  note      = {Open Access Book},
  url       = {https://institutionaldynamicsbook.culturalevolutionsociety.org/}
}

@book{weber2019economy,
  title={Economy and society: A new translation},
  author={Weber, Max},
  year={2019},
  publisher={Harvard University Press}
}

@article{katz2020complex,
  title={Complex societies and the growth of the law},
  author={Katz, Daniel Martin and Coupette, Corinna and Beckedorf, Janis and Hartung, Dirk},
  journal={Scientific Reports},
  volume={10},
  number={1},
  pages={18737},
  year={2020},
  publisher={Nature Publishing Group UK London}
}

@article{babu2004structure,
  title={Structure and evolution of transcriptional regulatory networks},
  author={Babu, M Madan and Luscombe, Nicholas M and Aravind, L and Gerstein, Mark and Teichmann, Sarah A},
  journal={Current Opinion in Structural Biology},
  volume={14},
  number={3},
  pages={283--291},
  year={2004},
  publisher={Elsevier}
}

@article{jeong2026dataset,
  title={{A Dataset Showing a Century of Evolution in the Complexity of the United States Legal Code}},
  author={Jeong, Dawoon and Holehouse, James and Yoon, Jisung and Kempes, Christopher P and West, Geoffrey B and Youn, Hyejin},
  journal={Scientific Data},
  year={2026},
  publisher={Nature Publishing Group UK London}
}

@article{Bentz2014-wz,
	author = {Bentz, Christian and Kiela, Douwe and Hill, Feli and Buttery, Paula},
	journal = {Corpus Ling. Ling.. Theory},
	month = jan,
	number = 2,
	title = {{Zipf's law and the grammar of languages: A quantitative study of Old and Modern English parallel texts}},
	volume = 10,
	year = 2014}

@article{berube2018lexdiv,
	author = {B{\'e}rub{\'e}, Nicolas AND Sainte-Marie, Maxime AND Mongeon, Philippe AND Larivi{\`e}re, Vincent},
	journal = {PLoS One},
	month = {07},
	number = {7},
	pages = {1-31},
	title = {Words by the tail: Assessing lexical diversity in scholarly titles using frequency-rank distribution tail fits},
	volume = {13},
	year = {2018}}

@book{juvenal_satires_vi,
  author    = {Juvenal},
  title     = {Satires},
  note      = {Satire VI, lines 347--348},
  year      = {ca. 2nd century CE}
}

@article{piantadosi2014zipf,
  title={Zipf’s word frequency law in natural language: A critical review and future directions},
  author={Piantadosi, Steven T},
  journal={Psychonomic Bulletin \& Review},
  volume={21},
  number={5},
  pages={1112--1130},
  year={2014},
  publisher={Springer}
}

@article{aldous2001stochastic,
  title={{Stochastic models and descriptive statistics for phylogenetic trees, from Yule to today}},
  author={Aldous, David J},
  journal={Statistical Science},
  pages={23--34},
  year={2001},
  publisher={JSTOR}
}

@article{Ferrer_i_Cancho2005-se,
	author = {Ferrer i Cancho, R},
	journal = {Eur. Phys. J. B},
	month = mar,
	number = 2,
	pages = {249--257},
	title = {{The variation of Zipf's law in human language}},
	volume = 44,
	year = 2005}

@article{yang2024leads,
  title={{What Leads to Administrative Bloat? A Dynamic Model of Administrative Cost and Waste}},
  author={Yang, Vicky Chuqiao and Grenier, Levi},
  journal={Proceedings of the National Academy of Sciences},
  note={{(\textit{In Press})}},
  year={2026}
}

@article{erwin2009evolution,
  title={The evolution of hierarchical gene regulatory networks},
  author={Erwin, Douglas H and Davidson, Eric H},
  journal={Nature Reviews Genetics},
  volume={10},
  number={2},
  pages={141--148},
  year={2009},
  publisher={Nature Publishing Group UK London}
}

@article{davidson2010emerging,
  title={Emerging properties of animal gene regulatory networks},
  author={Davidson, Eric H},
  journal={Nature},
  volume={468},
  number={7326},
  pages={911--920},
  year={2010},
  publisher={Nature Publishing Group UK London}
}

@incollection{mandelbrot1953informational,
  author    = {Mandelbrot, Benoit B.},
  title     = {{An Informational Theory of the Statistical Structure of Languages}},
  booktitle = {Communication Theory},
  editor    = {Jackson, W.},
  publisher = {Academic Press},
  address   = {Princeton},
  pages     = {486--502},
  year      = {1953}
}

@article{balasubrahmanyan1996,
	author = {Balasubrahmanyan, V. and Naranan, S.},
	journal = {Journal of Quantitative Linguistics},
	month = {12},
	pages = {177-228},
	title = {Quantitative Linguistics and Complex System Studies.},
	volume = {3},
	year = {1996}}

@book{zipf_psycho-biology_1935,
  address = {Oxford, England},
  author = {Zipf, George Kingsley},
  publisher = {Houghton Mifflin},
  shorttitle = {The psycho-biology of language},
  title = {The Psycho-Biology of Language: an Introduction to Dynamic Philology},
  year = 1935
}

@book{vonWright1968-VONAEI,
	address = {Amsterdam},
	editor = {Georg Henrik von Wright},
	publisher = {North-Holland Pub. Co.},
	title = {An Essay in Deontic Logic and the General Theory of Action: With a Bibliography of Deontic and Imperative Logic},
	year = {1968}
}

@article{Vamplew01072007,
	author = {Wray Vamplew},
	journal = {{The International Journal of the History of Sport}},
	number = {7},
	pages = {843--871},
	title = {Playing with the rules: Influences on the development of regulation in sport},
	volume = {24},
	year = {2007}}

@article{medina2005measuring,
  title={Measuring the evolution of transport protocols in the Internet},
  author={Medina, Alberto and Allman, Mark and Floyd, Sally},
  journal={ACM SIGCOMM Computer Communication Review},
  volume={35},
  number={2},
  pages={37--52},
  year={2005},
  publisher={ACM New York, NY, USA}
}

@article{hassel2008evolution,
  title={The evolution of a global labor governance regime},
  author={Hassel, Anke},
  journal={Governance},
  volume={21},
  number={2},
  pages={231--251},
  year={2008},
  publisher={Wiley Online Library}
}

@ARTICLE{Barabasi2004-bj,
  title     = "Network biology: understanding the cell's functional
               organization",
  author    = "Barab{\'a}si, Albert-L{\'a}szl{\'o} and Oltvai, Zolt{\'a}n N",
  journal   = "Nat. Rev. Genet.",
  publisher = "Springer Science and Business Media LLC",
  volume    =  5,
  number    =  2,
  pages     = "101--113",
  month     =  feb,
  year      =  2004,
  language  = "en"
}

@ARTICLE{Akaike-info-crit,
       author = {{Akaike}, H.},
        title = "{A New Look at the Statistical Model Identification}",
      journal = {IEEE Transactions on Automatic Control},
         year = 1974,
        month = jan,
       volume = {19},
        pages = {716-723},
          doi = {10.1109/TAC.1974.1100705},
       adsurl = {https://ui.adsabs.harvard.edu/abs/1974ITAC...19..716A}
}

@ARTICLE{Watts1998-nc,
  title     = {Collective dynamics of `small-world' networks},
  author    = "Watts, D J and Strogatz, S H",
  journal   = "Nature",
  publisher = "Springer Science and Business Media LLC",
  volume    =  393,
  number    =  6684,
  pages     = "440--442",
  month     =  jun,
  year      =  1998,
  language  = "en"
}

@article{vuong-test,
 ISSN = {00129682, 14680262},
 URL = {http://www.jstor.org/stable/1912557},
 author = {Quang H. Vuong},
 journal = {Econometrica},
 number = {2},
 pages = {307--333},
 publisher = {[Wiley, Econometric Society]},
 title = {Likelihood Ratio Tests for Model Selection and Non-Nested Hypotheses},
 urldate = {2026-04-07},
 volume = {57},
 year = {1989}
}

@article{PhysRevE.70.066111,
  title = {Finding community structure in very large networks},
  author = {Clauset, Aaron and Newman, M. E. J. and Moore, Cristopher},
  journal = {Physical Review E},
  volume = {70},
  issue = {6},
  pages = {066111},
  numpages = {6},
  year = {2004},
  month = {Dec},
  publisher = {American Physical Society},
  doi = {10.1103/PhysRevE.70.066111},
  url = {https://link.aps.org/doi/10.1103/PhysRevE.70.066111}
}

@article{Squartini2013-hi,
  title     = "Reciprocity of weighted networks",
  author    = "Squartini, Tiziano and Picciolo, Francesco and Ruzzenenti,
               Franco and Garlaschelli, Diego",
  journal   = "Sci. Rep.",
  publisher = "Springer Science and Business Media LLC",
  volume    =  3,
  number    =  1,
  pages     = "2729",
  year      =  2013,
  copyright = "https://creativecommons.org/licenses/by-nc-nd/3.0/",
  language  = "en"
}

@article{Jain2002-an,
  title     = "Large extinctions in an evolutionary model: the role of
               innovation and keystone species",
  author    = "Jain, Sanjay and Krishna, Sandeep",
  journal   = "Proceedings of the National Academy of Sciences",
  publisher = "Proceedings of the National Academy of Sciences",
  volume    =  99,
  number    =  4,
  pages     = "2055--2060",
  month     =  feb,
  year      =  2002,
  language  = "en"
}

@article{Bonacich2007-kv,
  title     = "Some unique properties of eigenvector centrality",
  author    = "Bonacich, Phillip",
  journal   = "Soc. Networks",
  publisher = "Elsevier BV",
  volume    =  29,
  number    =  4,
  pages     = "555--564",
  month     =  oct,
  year      =  2007,
  language  = "en"
}

@article{Taylor2017-xj,
  title     = "Eigenvector-based centrality measures for temporal networks",
  author    = "Taylor, Dane and Myers, Sean A and Clauset, Aaron and Porter,
               Mason A and Mucha, Peter J",
  journal   = "Multiscale Model. Simul.",
  publisher = "Society for Industrial \& Applied Mathematics (SIAM)",
  volume    =  15,
  number    =  1,
  pages     = "537--574",
  month     =  mar,
  year      =  2017,
  language  = "en"
}

@article{van2003scaling,
  title={Scaling laws in the functional content of genomes},
  author={van Nimwegen, Erik},
  journal={Trends in genetics},
  volume={19},
  number={9},
  pages={479--484},
  year={2003},
  publisher={Elsevier}
}

@online{Preamble,
   author = {Lord's},
   title={{Preamble to the Laws: Spirit of Cricket—The Laws of Cricket: Lord's}}, 
   year = 2025,
   note = {{\url{https://www.lords.org/mcc/the-laws/preamble-to-the-laws-spirit-of-cricket}}},
   urldate = {2026-05-16}
}

@article{coupette2021measuring,
  title={{Measuring law over time: a network analytical framework with an application to statutes and regulations in the United States and Germany}},
  author={Coupette, Corinna and Beckedorf, Janis and Hartung, Dirk and Bommarito, Michael and Katz, Daniel Martin},
  journal={Frontiers in Physics},
  volume={9},
  pages={658463},
  year={2021},
  publisher={Frontiers Media SA}
}

@article{bommarito2017measuring,
  title={{Measuring and modeling the US regulatory ecosystem}},
  author={Bommarito II, Michael J and Katz, Daniel Martin},
  journal={Journal of Statistical Physics},
  volume={168},
  number={5},
  pages={1125--1135},
  year={2017},
  publisher={Springer}
}

\clearpage
\appendix
\section*{Supplementary Materials}
\addcontentsline{toc}{section}{Supplementary Materials}

\renewcommand{\figurename}{Supplementary Figure}
\renewcommand{\thefigure}{S\arabic{figure}}
\setcounter{figure}{0}

\begin{figure}[ht]
    \includegraphics[width=0.7\textwidth]{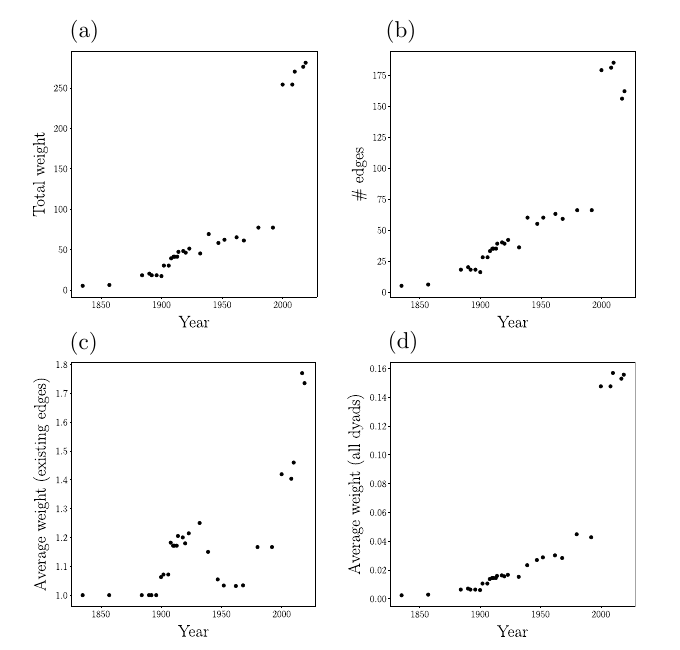}
    \caption{Descriptive statistics for the interdependency network representations of the cricket rule sets.}
    \label{fig:cit-ntwk-suppl-fig1}
\end{figure}

\begin{figure}[ht]
    \includegraphics[width=0.8\textwidth]{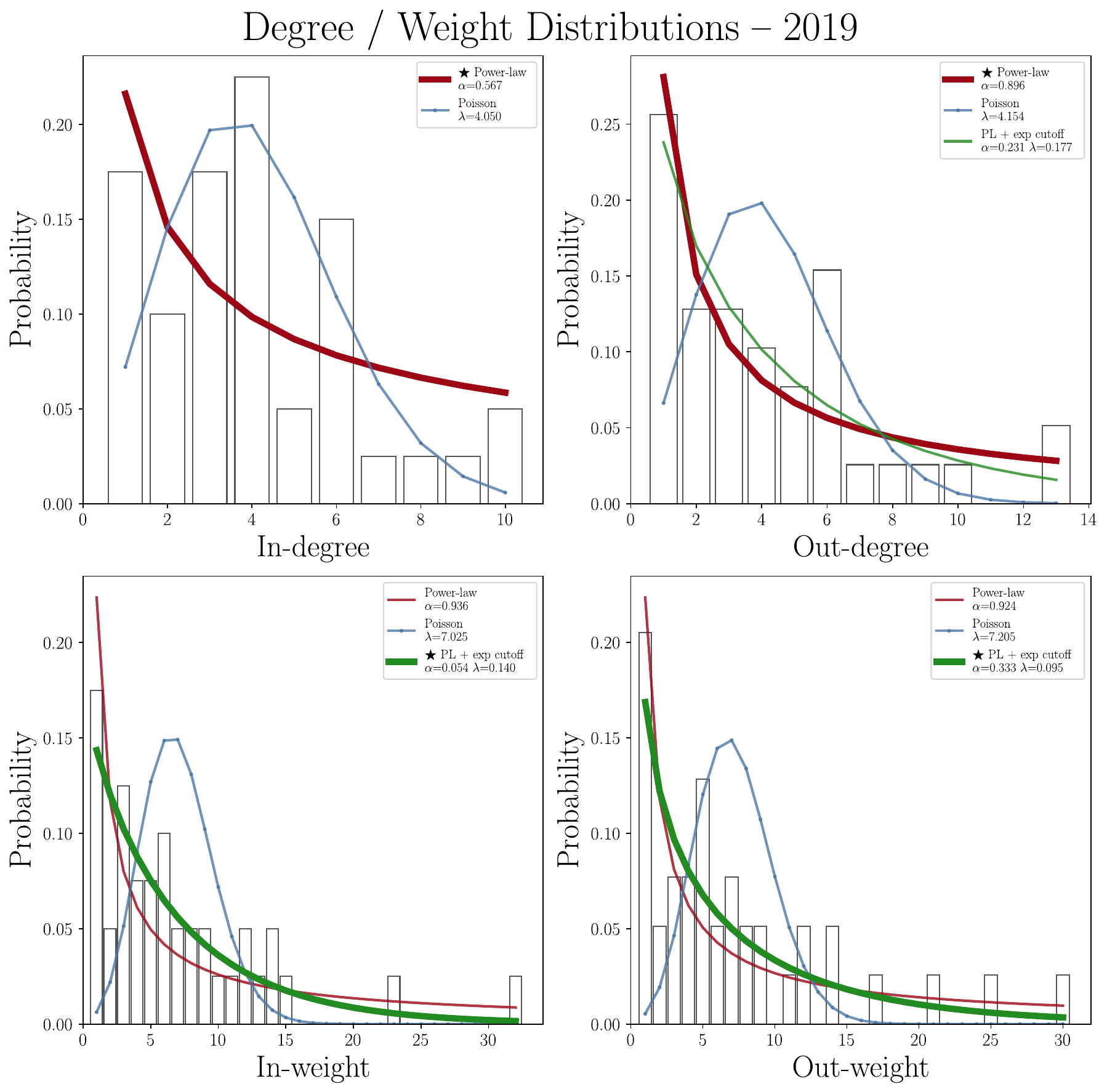}
    \caption{In-degree, out-degree, in-weight and out-weight distributions for the 2019 cricket rule set's interdependency network.}
    \label{fig:cit-ntwk-suppl-fig2}
\end{figure}

\begin{figure}[ht]
    \includegraphics[width=0.8\textwidth]{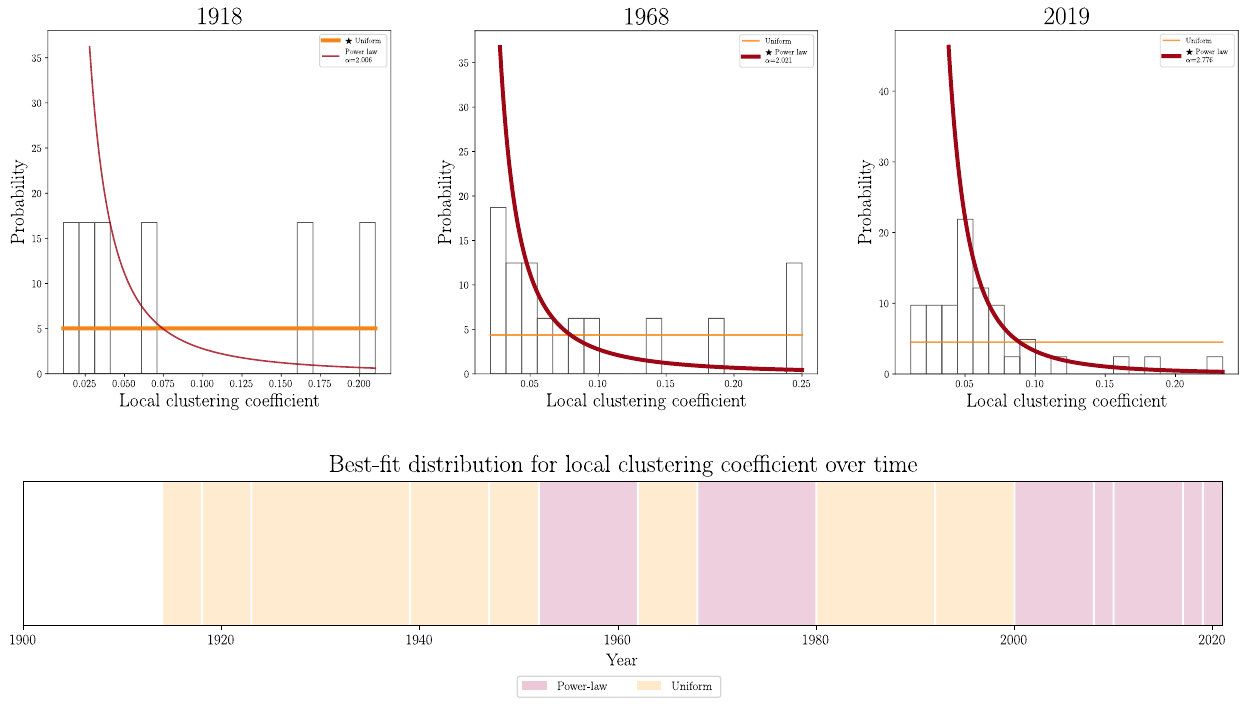}
    \caption{\textbf{Clustering coefficient distributions of the interdependency networks \cite{Watts1998-nc}.} Random and scale-free networks commonly exhibit a uniform clustering coefficient distribution, while hierarchical networks commonly exhibit a power law clustering coefficient distribution. Top row: clustering coefficient distributions for 1918, 1962 and 2019. Bottom: heatmap indicating the best model fit for clustering coefficient distribution over the duration of the rule set. We find consistent evidence that the power-law model is a better fit for the data from 2000 to 2019 \cite{vuong-test}. This transition suggests the emergence of a more hierarchical organization in the interdependency networks over time.}
    \label{fig:corr-coeff-suppl-fig}
\end{figure}




\begin{figure}[ht]
    \includegraphics[width=0.9\textwidth]{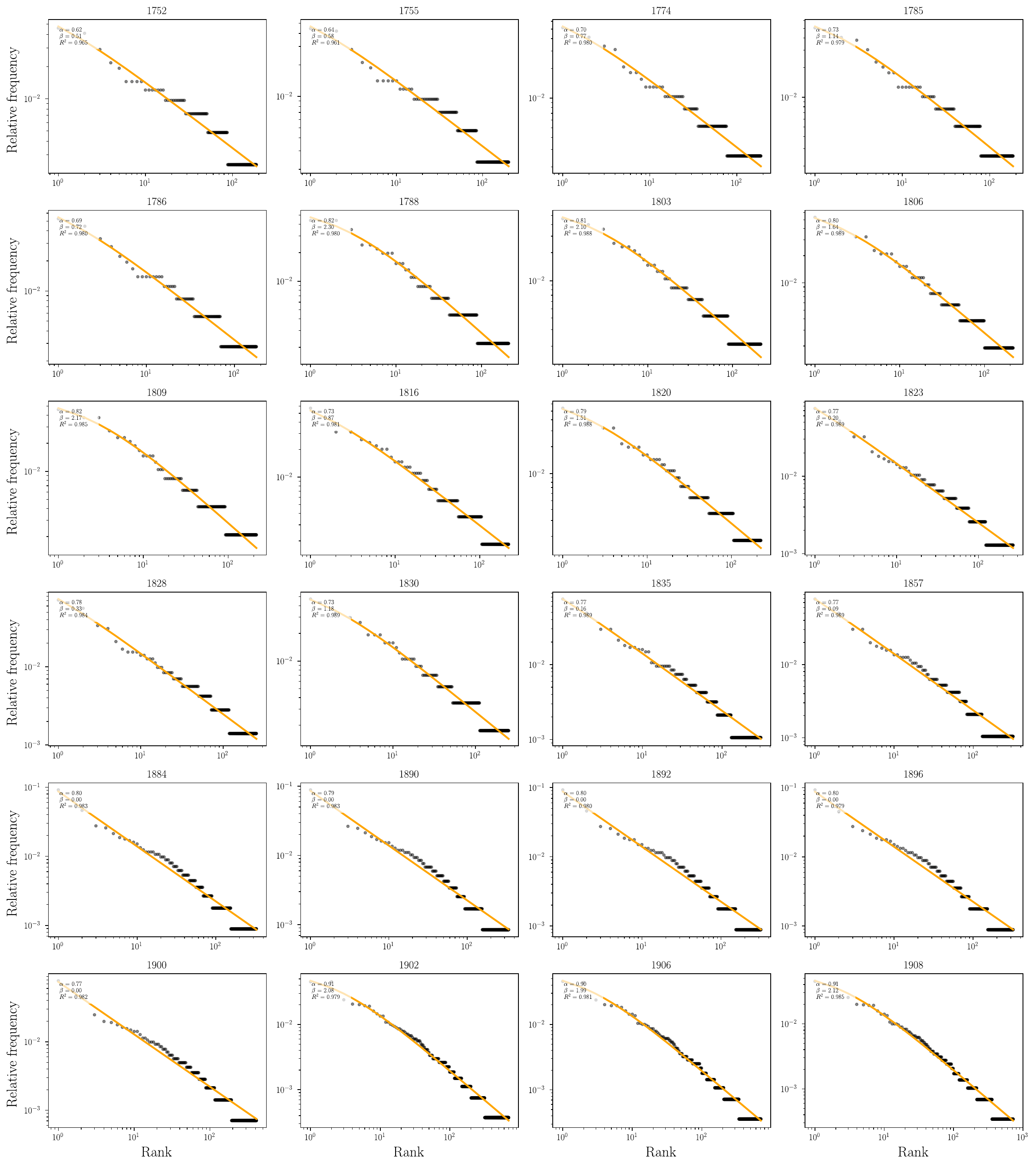}
    \caption{Zipf-Mandelbrot fits for the rank-frequency distributions of the first 24 rule sets.}
    \label{fig:all-zm-fits-p1}
\end{figure}

\begin{figure}[ht]
    \includegraphics[width=0.9\textwidth]{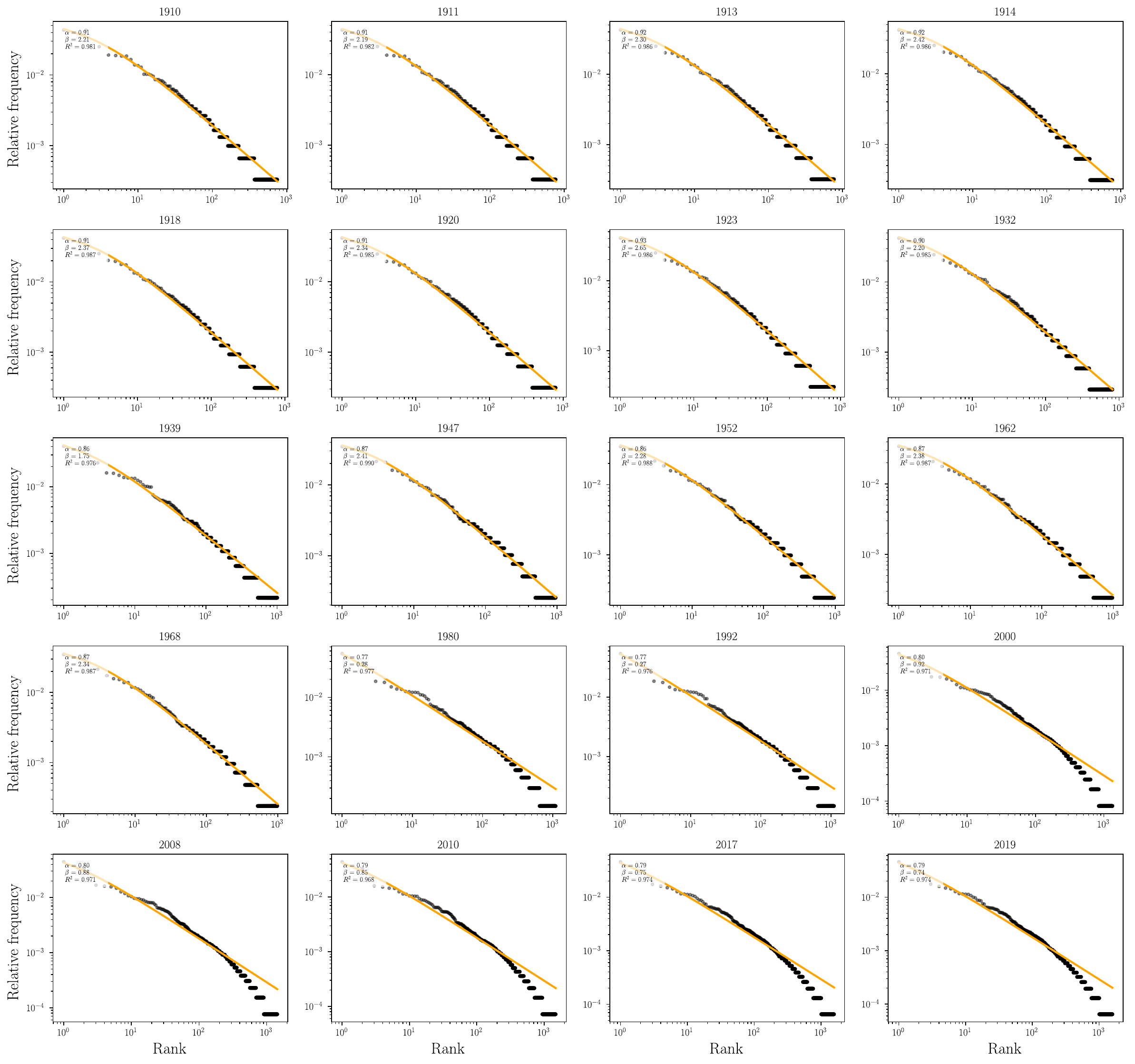}
    \caption{Zipf-Mandelbrot fits for the rank-frequency distributions of the last 20 rule sets.}
    \label{fig:all-zm-fits-p2}
\end{figure}

\begin{figure}[ht]
    \includegraphics[width=0.9\textwidth]{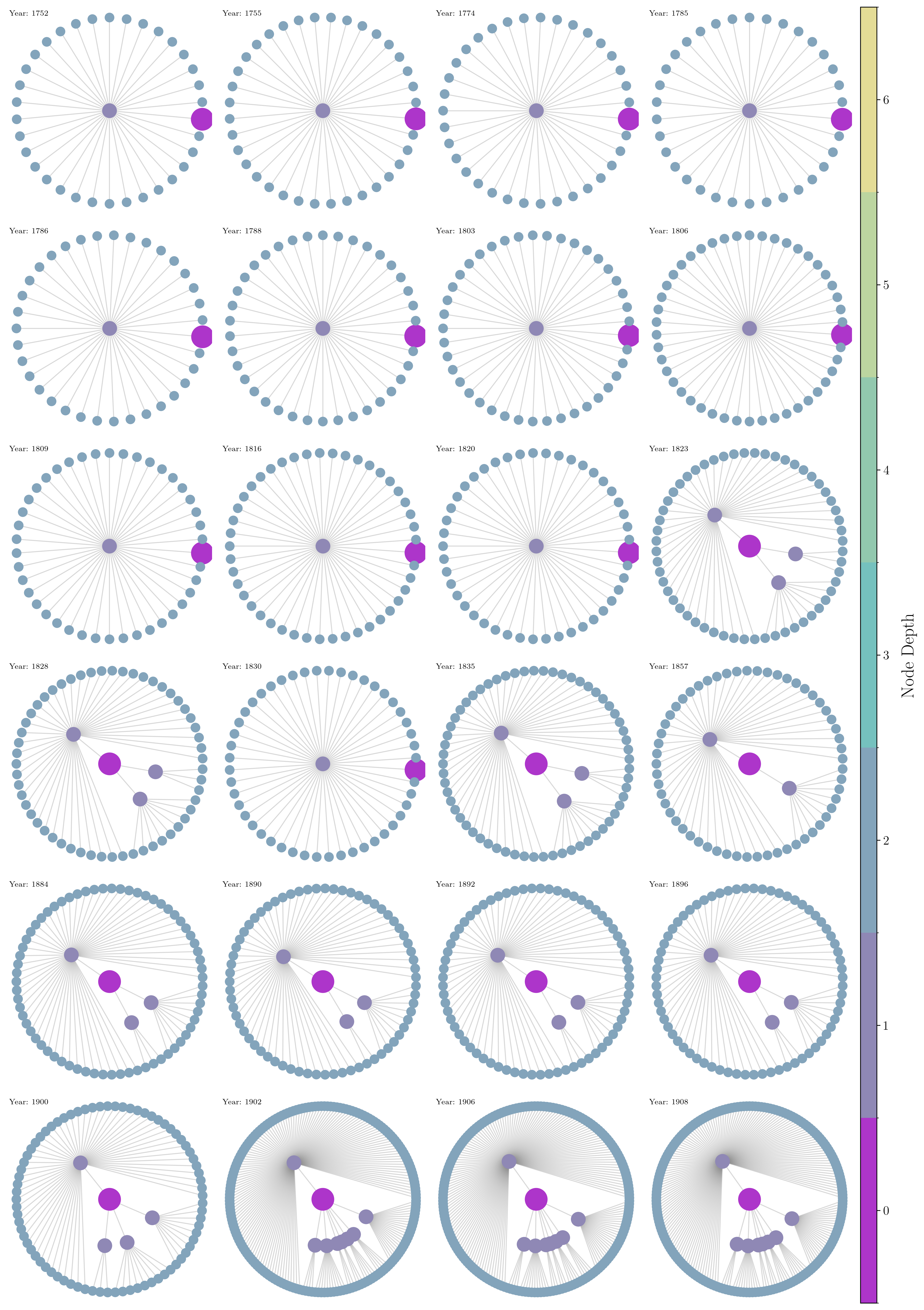}
    \caption{Rule trees over the first 24 rule sets.}
    \label{fig:all-years-ruleset-network-grid-p1}
\end{figure}

\begin{figure}[ht]
    \includegraphics[width=0.9\textwidth]{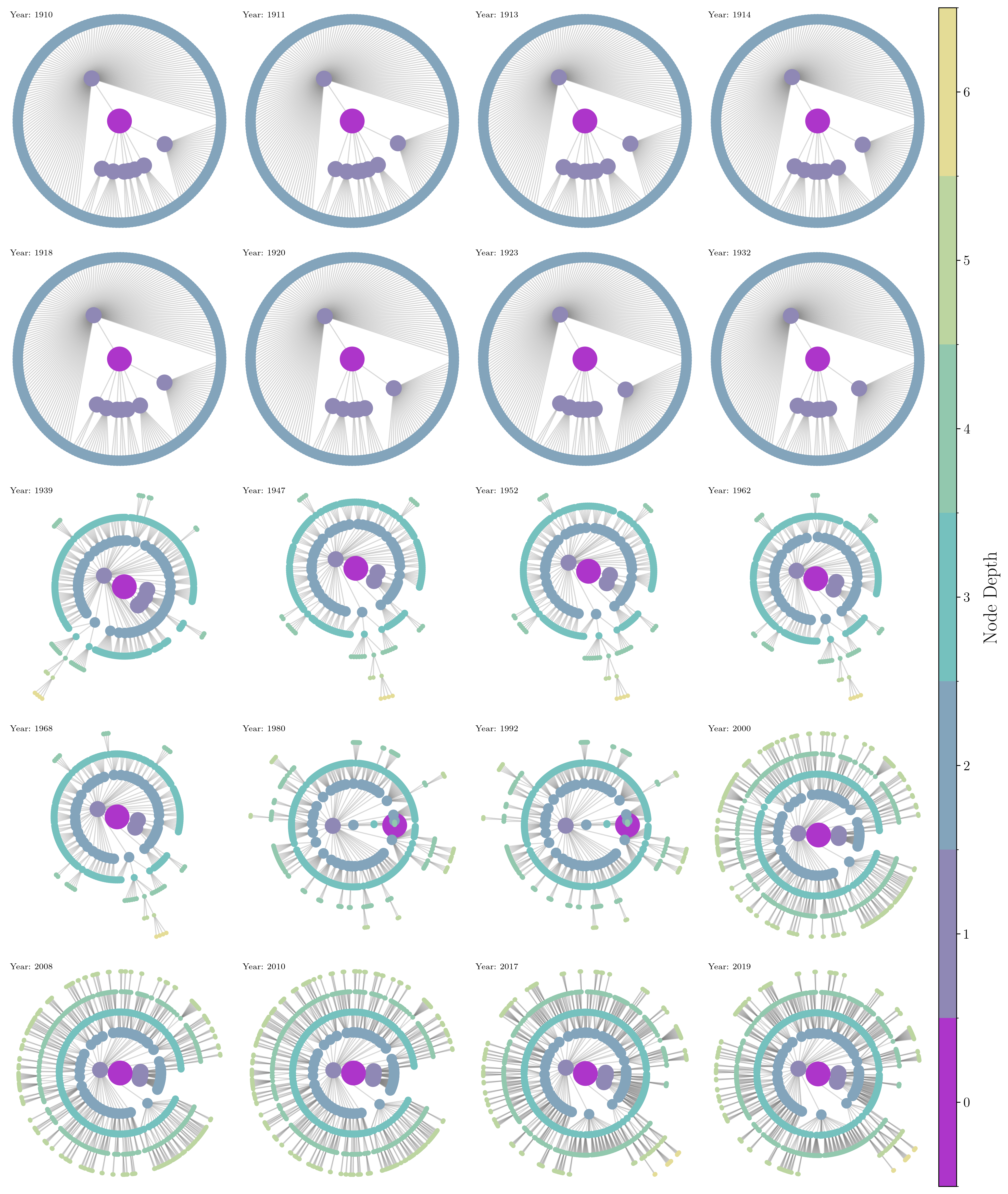}
    \caption{Rule trees over the last 20 rule sets.}
    \label{fig:all-years-ruleset-network-grid-p2}
\end{figure}

\begin{figure}[ht]
    \includegraphics[width=0.9\textwidth]{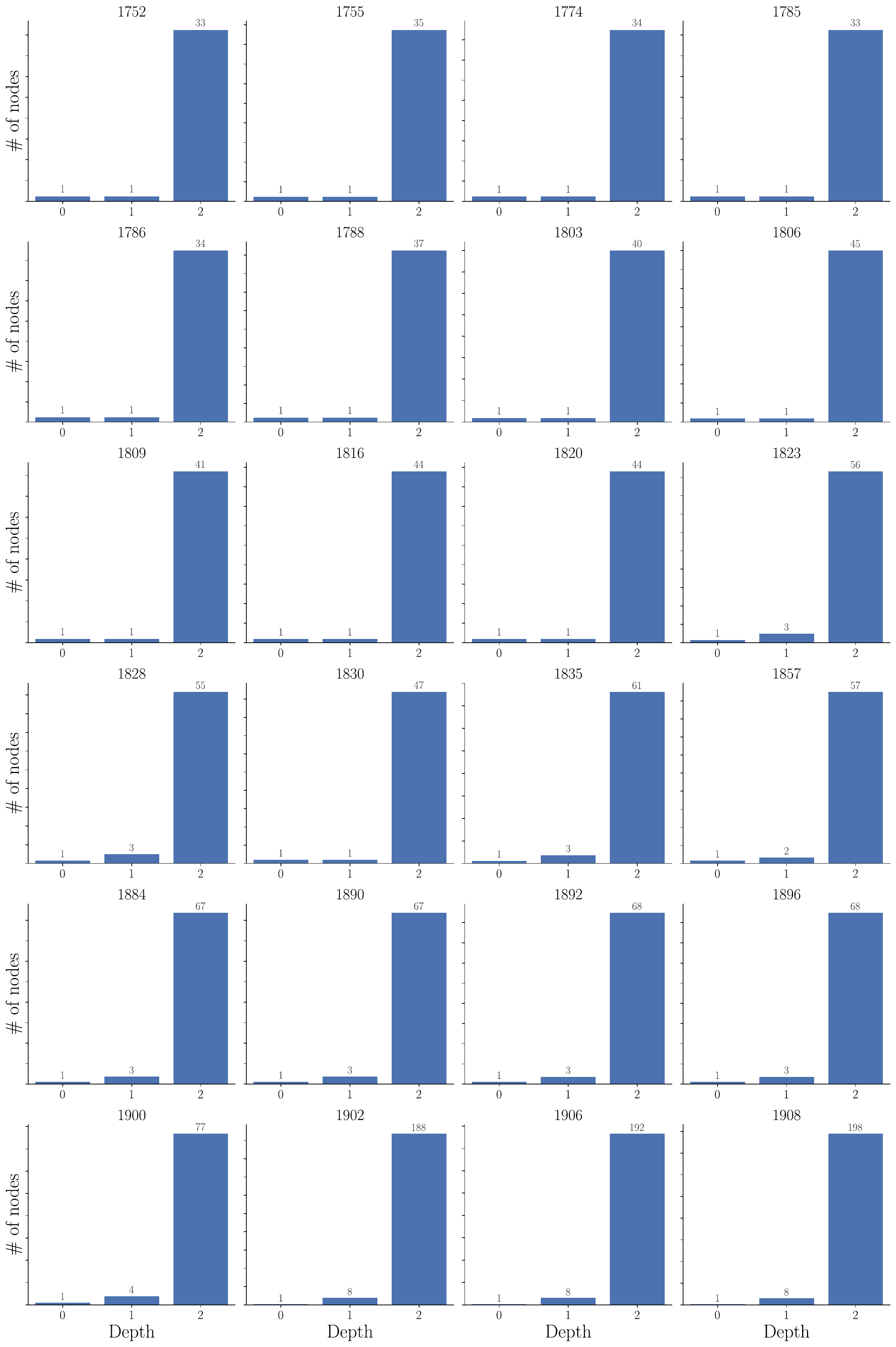}
    \caption{Depth distributions of the rule trees for the first 24 rule sets.}
    \label{fig:all-years-depth-distr-grid-p1}
\end{figure}

\begin{figure}[ht]
    \includegraphics[width=0.9\textwidth]{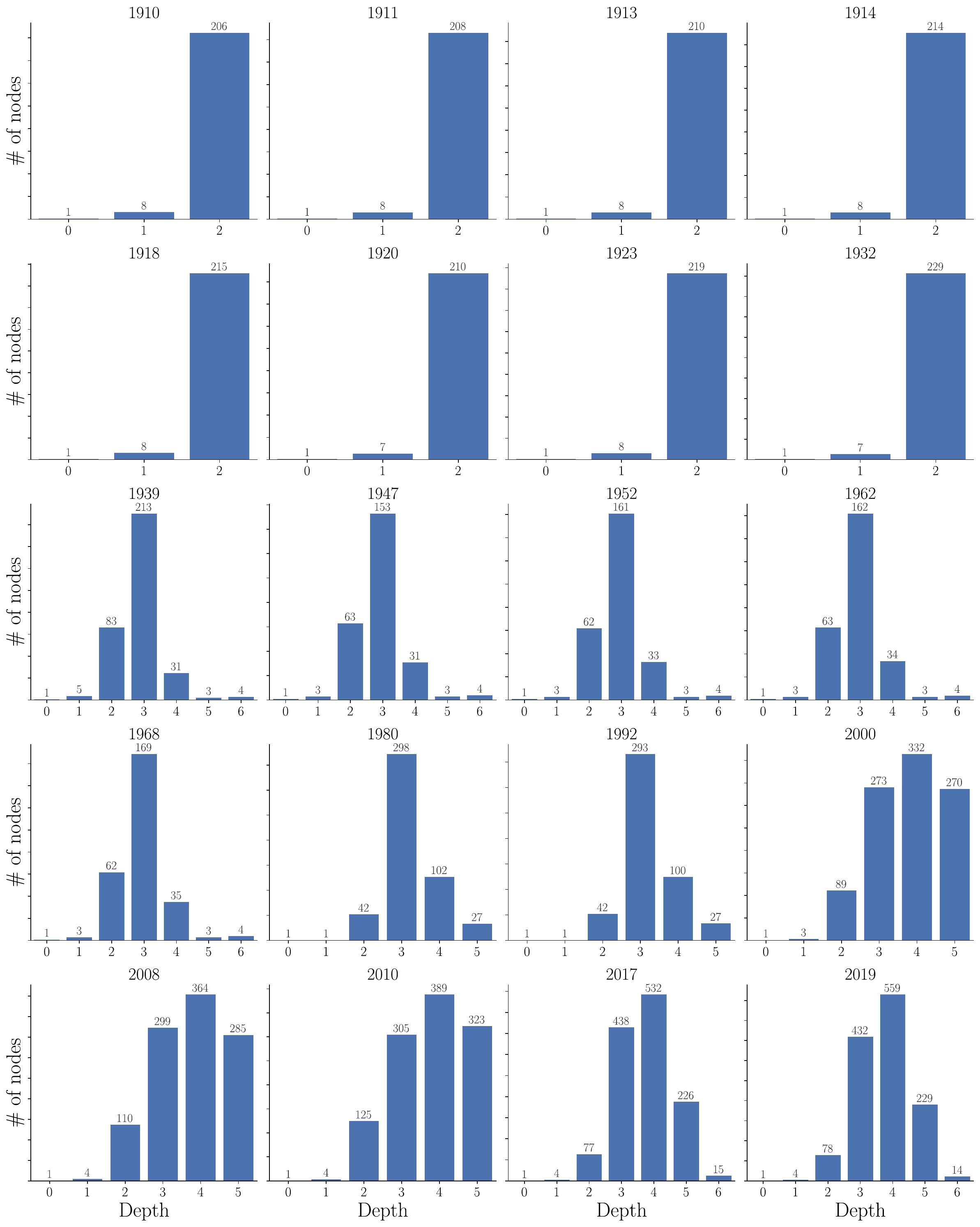}
    \caption{Depth distributions of the rule trees for the last 20 rule sets.}
    \label{fig:all-years-depth-distr-grid-p2}
\end{figure}

\end{document}